\documentclass[useAMS,usenatbib,onecolumn]{mn2e}
\usepackage{graphicx}
\usepackage{bm}
\usepackage{ulem}

\newcommand{\bea}{\begin{eqnarray}}
\newcommand{\eea}{\end{eqnarray}}
\newcommand{\beq}{\begin{equation}}
\newcommand{\eeq}{\end{equation}}

\newcommand{\simless}[0]{\mathbin{\lower 3pt\hbox
   {$\rlap{\raise 5pt\hbox{$\char'074$}}\mathchar"7218$}}}
\newcommand{\simgreat}[0]{\mathbin{\lower 3pt\hbox
   {$\rlap{\raise 5pt\hbox{$\char'076$}}\mathchar"7218$}}}

\newcommand{\lta}[0]{\simless}
\newcommand{\ra}[1]{\renewcommand{\arraystretch}{#1}}

\newcommand{\figref}[1]{figure \ref{#1}}

\newcommand{\capfigref}[1]{Figure \ref{#1}}
\newcommand{\capfigrefs}[1]{Figures \ref{#1}}
\newcommand{\eqnref}[1]{equation (\ref{#1})} %
\newcommand{\eqnrefs}[1]{equations (\ref{#1})} %
\newcommand{\eqnrefbare}[1]{(\ref{#1})} %
\newcommand{\capeqnref}[1]{Equation (\ref{#1})}

\graphicspath{{}{figures/}}

\title[Non-linear evolution using LPT re-expansions]{Modelling non-linear evolution using Lagrangian Perturbation Theory (LPT) re-expansions}

\author[Sharvari Nadkarni-Ghosh and David F. Chernoff]{Sharvari Nadkarni-Ghosh$^{1,}$$^{2}$\thanks{E-mail:
sharvari@iitk.ac.in} and David F. Chernoff$^{3}$\thanks{E-mail:
chernoff@astro.cornell.edu} \\
$^{1}$Department of Theoretical Sciences, S.N. Bose N.C.B.S, Sector III, Block JD, Salt Lake, Kolkata, W.B. 700098 India\\
$^{2}$Department of Physics, I.I.T. Kanpur, Kanpur, U.P. 208016 India \\
$^{3}$Department of Astronomy, Cornell University, Ithaca, NY 14853 USA}

\begin{document}
\date{}
\volume{}
\pagerange{\pageref{firstpage}--\pageref{lastpage}} \pubyear{2013}

\maketitle

\begin{abstract}

We present a new method to calculate formation of
cosmological structure in the Newtonian limit. The method
is based on Lagrangian perturbation theory plus two key
theoretical extensions. One advance involves identifying and
fixing a previously ignored gauge-like degree of freedom
relating quantities calculated in LPT to those measured
by a preferred Friedmann-Robertson-Walker (FRW)
observer. Handling this connection between calculational and observer
frames is physically essential and ensures a momentum
conserving description.  The second extension is to systematically
re-expand the equations of motion to increase LPT's radius of
convergence to the maximum future time prior to orbit
crossing.  The paper implements a complete algorithm and performs
extensive ``proof of principle'' tests of the new method,
including direct comparison to known solutions, evaluation
of conserved quantities and formal convergence studies. All
are satisfactory. We show convergence is exponential in grid size
and Lagrangian order and polynomial in step size. There are
three {\it powerful advantages} afforded by the new technique:
(1) it employs a smooth representation of all fields and the
results are not limited by particle induced shot-noise
errors, (2) it permits the numerical error to be controlled
by changing Lagrangian order and/or number of steps
allowing, in principle, arbitrarily small errors to be
achieved prior to orbit crossing and (3) it handles generic
cold initial data (any periodic density and velocity fields,
including those with initial rotational components).
Together, these properties make the new technique
well-suited to handle quasi-linear scales where analytic
methods and/or numerical simulations fail to provide
suitably accurate answers.

\end{abstract}

\begin{keywords}
cosmology: theory -- large-scale structure of Universe. 
\end{keywords}

\section{Introduction}
An important means of determining and/or constraining
cosmological parameters is to compare the growth of
structure as predicted by theoretical models with that
observed in the actual universe. The complexity and
sophistication of theoretical approaches span an enormous
range of possibilities.  Linear perturbation theory provides
the simplest analytic
description of growth, generally suitable when the perturbation expansion
parameter is small.  Quasi-linear theory extends the
accuracy and applicability of the description at the cost of
increasingly complex formulations. Finally, an N-body
simulation is capable of handling fully non-linear situations and 
delivering, in principle, exact answers albeit at substantial
computational cost and without the benefit of analytic
insight.  The choice of a
particular method is best determined by the sort of problem
one intends to solve. For example, if one is interested in
calculating how an initial Gaussian perturbation spectrum
generates non-Gaussianity as structure forms
then one seeks an accurate method for treating perturbations
once mode-mode coupling becomes important. As in Goldilocks's
fable the extreme choices may prove unpalatable: on the one hand, 
the coupling of interest is absent in a purely linear analysis; on the other,
particle-based methodologies introduce spurious shot
noise in representations of the underlying smooth density
distribution so that high accuracy may demand infeasible
numbers of particles.

This paper's focus is the development of a middle
approach, a method for the numerical treatment of structure
formation based on intrinsically smooth descriptions of the
density and velocity fields and capable of tracing
quasi-linear evolution to arbitrary
accuracy. We anticipate it will find application in
situations where small non-linear effects and/or highly
accurate results prior to collapse are of primary interest. The
physical context is Newtonian cosmology i.e. sub-horizon
scales, non-relativistic velocities and
weak gravitational fields.  The methods are
applicable until the formation of the first caustic.  Future
studies will extend the methodology to reach beyond
particle crossing and to handle physical scales
approaching that of the horizon.

Lagrangian perturbation theory (LPT), the heart of
the approach employed here, has been well-studied. Lagrange variables in the context of gravity were introduced more than four decades ago  (\citealt{novikov_nonlinear_1969}). Its use
in cosmology was initiated by Zel'dovich who gave an
approximate analytic solution for
special initial conditions with peculiar velocity proportional to 
peculiar acceleration
\citep{zeldovich_gravitational_1970}.
This is the famous ``Zel'dovich approximation'' for the evolution
of growing modes where
``approximate'' means accurate to first order in particle displacements.
Buchert's studies (\citealt{buchert_class_1989}; \citealt{buchert_lagrangian_1992}, henceforth B92)
enlarged the class of initial conditions that could be
solved to first-order. Generalizations of the equations of motion, the
initial conditions, the perturbation order and the method of solution
have appeared regularly 
(\citealt{moutarde_precollapse_1991}; \citealt{bouchet_weakly_1992}; 
\citealt{buchert_lagrangian_1993}, henceforth BE93; \citealt{buchert_lagrangian_1994}, henceforth B94; \citealt{catelan_lagrangian_1995}; \citealt*{munshi_nonlinear_1994}; \citealt{bouchet_perturbative_1995}; \citealt{ehlers_newtonian_1997}, henceforth EB97; \citealt{rampf_lagrangian_2012};  \citealt{rampf_zeldovich_2012}; \citealt{rampf_lagrangian_2012-1}; \citealt{rampf_recursion_2012}; \citealt{tatekawa_fourth-order_2012}). Physical extensions include general relativistic equations of motion
\citep{kasai_tetrad-based_1995} and the treatment of relativistic fluids (\citealt{matarrese_post-newtonian_1996}; \citealt*{matarrese_general-relativistic_1993, matarrese_relativistic_1994}).

LPT has proven to be a practical tool in many different
applications including modelling the non-linear halo mass
functions (\citealt{monaco_lagrangian_1997}; \citealt*{monaco_pinocchio_2002}; \citealt{scoccimarro_pthalos:_2002}),
reconstructing baryon acoustic oscillations (\citealt{eisenstein_improving_2007}), modelling the non-linear density velocity relation (\citealt{kitaura_estimating_2012}) and setting up the 
initial conditions for fully numerical
simulations (\citealt{scoccimarro_transients_1998}; \citealt*{crocce_transients_2006}).

\begin{figure}
\includegraphics[height=6cm]{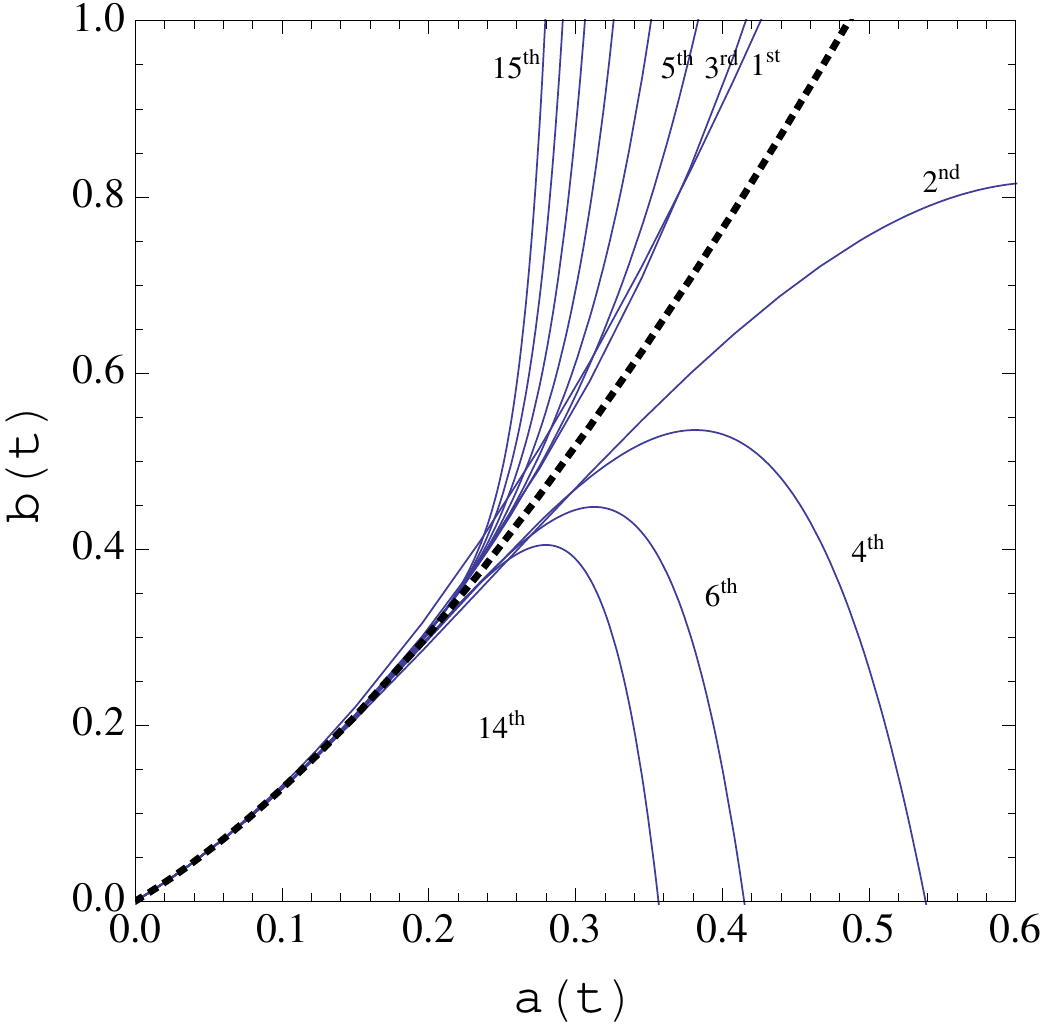}
\caption{Convergence of LPT: $a(t)$ and $b(t)$ are the scale
  factors of the background and the underdense spherical
  perturbation, respectively. The black dotted line is the
  exact open model with $\delta \rho/\rho = -9 \times
  10^{-3}$ at $a=10^{-3}$ and fractional Hubble velocity $3
  \times 10^{-3}$.  The blue lines show successive higher
  order approximations from $1$ to $15$ (as labeled).
  The LPT series converges only $a \lta 0.2$ even
  though there is no future singularity in this
  cosmology. LPT re-expansion provides a method to extend
  the solution to arbitrarily large $a$ in this example. }
\label{openplot}
\end{figure}

An assumption common to all these treatments is that the
matter velocity at a point in space is single valued (the
distribution is ``cold''). LPT breaks down once particle
orbits begin to cross, caustics appear, densities diverge
and physical singularities first form. Nothing in this paper
ameliorates the onset of these effects. There have been
interesting suggestions on how one might modify LPT to handle this
fundamental limitation (e.g. \citealt{adler_lagrangian_1999}) 
but we do not pursue the issue here.
For our purposes the immediate concern is a new
limitation of LPT that has recently come to light. The
convergence of the formal series expansion which is the
basis of LPT turns out to be limited in time (\citealt{nadkarni-ghosh_extending_2011},
hereafter NC11). No one would be surprised that the onset of
particle crossings inhibits the utility of the formal
perturbation expansion for subsequent times. What was unexpected and striking to us was
the existence of limits in a variety of other situations,
even cosmologies lacking future particle crossings and
devoid of future physical singularities. To
the best of our knowledge 
\citet{sahni_accuracy_1996} were the first to
report the failure of LPT
to describe one of the simplest cosmological problems, the
evolution of spherical homogeneous voids. We confirmed their
result, investigated the issue for all top-hat cosmologies
and traced the problem to the occurrence of poles in a
suitable set of analytically continued equations of motion.
We complemented this mathematical explanation with direct
numerical evaluation of the LPT series in all qualitatively
distinct regimes for top-hat like cosmologies.  The upshot is the existence of a
non-trivial limit to the future times for which the LPT
description converges. We dubbed the interval when
convergence of the perturbation expansion occurs
as the ``time of validity.'' Even if one could work to infinite
order in the perturbation amplitude the LPT series would
fail to give the correct answer at times outside the interval.
The end of the time of validity {\it precedes} any limitation due
to particle crossing. \capfigref{openplot} shows
the first 15 orders of the LPT description for a
simple underdense top-hat. The time of validity extends only up until
$a \sim 0.2$.

Like us many readers may find it odd that this behaviour had not been previously
reported in a model as well-studied as the top-hat but we can
point to two reasons. First,
convergence of LPT had never been addressed in a systematic
fashion. Second, for collapsing Zel'dovich initial conditions
(initial velocity and acceleration proportional) the situation is degenerate in the following
sense: the convergence limitations become identical to the moment of
caustic formation. Generically, however, the time
of validity for top-hat evolution does not extend up to the first particle
crossings and is the more stringent restriction.
To calculate the
``answer'' beyond the time of validity we proposed to evaluate the
solution at an intermediate time when the series is still
valid and use the results as initial conditions for a new
expansion about the intermediate point. We worked out the
details for this solution technique which we dubbed ``LPT
re-expansion''. It is roughly analogous to analytic
continuation of a power series in the complex plane.
Re-expansion will not circumvent the ultimate limitation set
by particle crossing but it will allow the
solution to be extended to this maximum possible future time
after which the flow is no longer cold. We provided
detailed numerical examples of how LPT re-expansion handles
the previous difficulties encountered in the top-hat problem.

Although our analysis was specific to one of the simplest of all
cosmological problems we believe these ideas are valid in
more general circumstances. Consequently,
one should shift from thinking of LPT as a one step, stand-alone method of calculation
to viewing it as the fundamental operation
in a numerical, multi-step method of
solution. Our development of LPT re-expansion resembles a traditional
finite-difference method in that the system is updated on a
step-by-step basis. Just as is true for particle-based
calculations, stability limits how big a step can be
taken and accuracy varies with step size and expansion
order.

This paper works out LPT with re-expansion for general
inhomogeneous initial conditions in a flat
Friedmann-Robertson-Walker (FRW) background cosmology.  We present
fundamental mathematical results, practical computational
methods and stringent numerical tests.  This paper serves as
a `proof of principle' demonstration of the method. We
will provide practical cosmological applications in
subsequent papers.

In \S \ref{sec:setup} we describe the Lagrangian framework
and the relationship of the full solution of
the geodesic equation of motion for particles to that
provided by the LPT expansion.  The formal expression for
the latter is derived by first taking spatial
derivatives of the geodesic equation (B92; BE93; B94; EB97). This system is invariant under spatially uniform, time-dependent translations. 
As we show in some
detail, it is necessary to fix this degree of freedom to recover the physical solution. This involves supplementing the order-by-order LPT solution with purely temporal functions, which we refer to as `frame shifts'. This issue has never been addressed
for single-step applications of LPT. For some initial
conditions the omission turns out to be inconsequential
but for most others
it breaks momentum conservation. Moreover, when multiple steps are
taken it becomes crucial to employ the full solution at each
stage of the calculation. We will show that the numerical method
converges if and only if that is done.

This paper notably differs from most previous approaches by
considering initial conditions of a completely general nature.  LPT is
often used to propagate the small fluctuations present at
recombination to a modest redshift for the purpose of
initialising an N-body treatment. In studies of cosmological
structure formation it is common to begin from Zel'dovich
initial conditions since vorticity modes decay away in an
expanding universe. The damping is normally sufficient to
obviate the influence of the initial vorticity so that an LPT
treatment of the growing modes alone is adequate for setting
up an N-body treatment of structure formation in pure $\Lambda$CDM.
Vorticity is intrinsically generated once non-linearities form
and particle crossing begins; these complications are accurately
handled by traditional N-body methods. In this context,
there is little need for treating arbitrary initial data.

Of course having a capability to study both longitudinal and transverse
motions opens up interesting possibilities involving
additional participants in structure formation such as
magnetic fields and cosmic strings since these may act to
introduce vorticity into the cosmic flows at late times. But the main
motivation to work in full generality is to facilitate the
use of LPT as part of a future calculational approach meant to
treat non-linear structure formation.  In any
realistic multi-step numerical approach (analogous to that
developed herein) vorticity will inevitably appear just as
it does in traditional N-body treatments. After it arises each
step of a multi-step calculation must be able to handle
essentially arbitrary density and velocity configurations as
initial data. The ability to treat these initial conditions
with LPT is a pre-requisite for utilizing LPT itself in the
future approach. That is the main reason we work in full generality.

\S \ref{sec:setup} includes an outline of the design and
implementation of the re-expansion algorithm. \S \ref{sec:tests}
summarizes tests for both special and general initial conditions.
We test the algorithm's
ability to reproduce known solutions. We analyse the dependence of
errors on Lagrangian order, step size and size of the
numerical grid. Taken together these tests validate the
concept of Lagrangian re-expansion, its analytic form and
our computational implementation. All tests are for small
grid sizes but we expect the methods demonstrated to apply
without essential modification to larger scale calculations.

\S \ref{sec:conclusion} presents conclusions and possible
future developments.

\section{Gravitational field equations in the Lagrangian framework}  
\label{sec:setup}
\subsection{General Setup}

 \label{sec:setupA}
\def\ainit{{a_0}} 
\def\Hinit{{H_0}}
\def\mattersubscript{{m}}
\def\rhominit{{\rho_{\mattersubscript,0}}} 
\def\rhomav{{\rho_{\mattersubscript}(t)}} 
\def\rhom{{\rho_{\mattersubscript}({\bf r},t)}} 
\def\rhomtzero{{\rho_{\mattersubscript}({\bf r},t_0)}} 
\def\rhomofx{{\rho_{\mattersubscript}({\bf x},t)}} 
\def\rhomofxOBS{{\rho_{\mattersubscript}({\bf x}_{OBS},t)}} 
\def\rhomofX{{\rho_{\mattersubscript}({\bf X},t)}} 
\def\rhomofXtzero{{\rho_{\mattersubscript}({\bf X},t_0)}} 
\def\rhomofY{{\rho_{\mattersubscript}({\bf Y},t)}} 
\def\rhomofYtzero{{\rho_{\mattersubscript}({\bf Y},t_0)}} 
\def\darkenergysubscript{{X}}
\def\rhoXinit{{\rho_{\darkenergysubscript,0}}}
\def\rhoXav{{\rho_{\darkenergysubscript}(t)}} 
\def\OmegaXinit{{\Omega_{\darkenergysubscript,0}}}

Assume a flat FRW background
cosmology with scale factor $a(t)$ that contains
pressureless dark matter with average
density $\rhomav$ and time-dependent dark energy with
average density $\rhoXav$ having fixed equation of state
parameter $w$.  We do not distinguish between cold baryons
and cold collisionless matter and we work before the
formation of caustics.  The homogeneous background evolution
is determined by the initial scale factor $\ainit$, Hubble
constant $\Hinit$ and matter and dark energy densities,
$\rhominit$ and $\rhoXinit$ respectively. The subscript
``0'' indicates time $t_0$ which may differ from the current
epoch.

We intend that the initial conditions allow for arbitrary
density and velocity perturbations of cold dark matter
subject to minimal restrictions.  We constrain the dark
energy component to be spatially uniform at all times \footnote{Although it has been suggested that quintessence
  models with $w \neq -1$ should have spatial fluctuations
  (for example
  \citealt*{caldwell_cosmological_1997}), these fluctuations are
  estimated to be small (for example \citealt*{mota_magnitude_2008};
  \citealt*{cooray_measuring_2010}).} .  
We work in an expanding,
periodic volume, a 3-torus. For physical, cosmological applications
this volume is intended to be (1) a fair representation of
the actual universe and (2) one that can be accurately
treated in the Newtonian limit.  Our perturbative approach does not allow
any back-reaction of the small-scale physics on the
large-scale expansion of the homogeneous and isotropic background. Consistency
requires that the initial mean density of the inhomogeneous
matter distribution agree exactly with that of the
background cosmology. Mathematically, the evolution of
the inhomogeneous Newtonian problem is completely determined by the
otherwise arbitrary, initial density and velocity perturbations. 
We must check, in principle, that the Newtonian
limit is valid at all times.

\subsection{Lagrangian Picture}
We will start by assuming that the coordinate system of the
periodic volume coincides with one of the family of
preferred world lines in FRW cosmology, i.e. one for which
the CMB dipole vanishes. We will refer to this starting point as
the OBS ``inertial frame'' and the observers as ``inertial
observers''.

In the Lagrangian picture the particle position is a function of time and a
set of Lagrangian labels ${\bf Y}$.  The labels are assigned
at the initial time and, for our purposes, are
always taken to be initial comoving Eulerian positions.  We write
${\bf x}(t; {\bf Y})$ for the instantaneous comoving
coordinate of a fluid particle.  By contrast, in the
Eulerian picture ${\bf x}$ is an independent variable that
identifies a fixed grid position. For notational simplicity
we will often use the descriptive shorthand that ${\bf x}$
is either Lagrangian (equivalently, a particle) or Eulerian.

In many applications one starts with a finite, discrete set of
particles (e.g. an N-body code) but here we assume space is
completely filled with particles. In effect, we promote the
discrete set of Lagrangian labels to a continuous field throughout space so that
at any time it is possible to relate an Eulerian grid
coordinate to a Lagrangian label in a one-to-one
fashion. Mathematically, the existence of a well-defined
velocity field at every point in space establishes this
result (EB97).  The Lagrangian
particle velocity is identical to the Eulerian velocity at
the grid position that coincides with the particle position.

Throughout this paper we will use the following convention
for density: $\rho({\bf r},t)$ is the physical density
expressed as a function of ${\bf r}$, the physical position,
and $\rho({\bf x},t)$ is the same numerical quantity
expressed as a function of ${\bf x}$,
the comoving coordinate. Explicitly, we have
\bea
\rho({\bf x},t) & \equiv & \rho({\bf r},t)|_{{\bf r}=a{\bf x}} \\
\Delta M        = \rho({\bf r},t) d^3r 
                & = & \rho({\bf x},t) d^3r = \rho({\bf x},t) a^3 d^3x  .
\eea
Note that $\rho({\bf x},t)$ is {\it not} the comoving density.

Likewise, we employ a similar convention for density written in
terms of the Lagrangian variable
\bea
\rho({\bf Y}, t) & \equiv & \rho({\bf r},t)|_{{\bf r} = {\bf r}({\bf Y},t)} \\
\Delta M         = \rho({\bf r},t) d^3r 
                 & = & \rho({\bf Y},t) d^3r = 
                      \rho({\bf Y},t) J({\bf Y},t) d^3Y \\
J({\bf Y},t) & = & {\rm Det} \left( \frac{\partial {\bf r}}{\partial {\bf Y}} \right)
\eea
where ${\bf r}({\bf Y},t)$ is the physical position of the
particle labeled by Lagrangian coordinate ${\bf Y}$ and $J$
is the determinant of the coordinate transformation. The distinguishing
feature of the Lagrangian labels is that they
track the mass so, in addition to the above general properties, 
we also have mass conservation in the form
\beq
\rho({\bf Y},t) J({\bf Y},t) =  \rho({\bf Y},t') J({\bf Y},t') 
\label{eq:rhoy}
\eeq
for any times $t$ and $t'$.

In comoving coordinates the Newtonian limit of the
geodesic equation and of Poisson's equation for the
gravitational potential $\psi({\bf x},t)$ are
\bea 
\frac{d}{dt} \left( a^2
\frac{d {\bf x}}{dt} \right) &=& -{{\bf \nabla}}_{x}
\psi ({\bf x}, t)
\label{originaleq} \\
\nabla_x^2 \psi ({\bf x}, t)&=& 4 \pi G a^2 \delta \rhomofx 
\label{originaleqII}
\eea
where $a(t)$ is the scale factor and $\delta \rhomofx$ is the physical matter density
perturbation. There is also an equation for mass conservation. 
The $x$-related terms are Eulerian except
for those on the left-hand side of the first equation which
refer to particles ($d/dt$ and ${\bf x}$).  
The time-derivative is the `Lagrangian' or `convective' derivative,
\beq 
\frac{d}{dt} =  \frac{\partial}{\partial t} + {\bf v} \cdot \nabla_{x},   
\eeq
where ${\bf v}$ is the Eulerian velocity. The operator
commutes with the Lagrangian spatial derivative but not the Eulerian one.

To make mathematical sense of these
equations as a Lagrangian system we should, schematically, specify Lagrangian
labels at the initial time and transform the Eulerian to
Lagrangian forms while imposing the condition that the
Eulerian coordinate match the particle position. 

We regard the system as posing an initial value problem
for the particle positions and velocities. If the initial
conditions are specified as Eulerian density and velocity
perturbations then they must be transformed to particle
positions and velocities.  The density perturbation is
a functional of the initial density and the
initial and final particle positions. The potential
can be solved at any time the density is known.

In physical coordinates ${\bf  r} = a(t) {\bf x}$ we have
\bea
\label{originaleq2A} 
a {\ddot {\bf r} } - {\ddot a} {\bf r} &=&- a {\bf \nabla}_r \psi({\bf r},t) \\
\label{originaleq2B} 
\nabla^2_r \psi({\bf r},t) &=&   4 \pi G  \delta \rhom ,
\eea
where the dot denotes the total derivative with respect to time. As
above, ${\bf r}(t)$ and $\ddot {\bf r}(t)$ (left-hand side of the
first equation) refers to particles while
the structure of the remaining terms is Eulerian. 

Buchert and Ehlers' derivation of the Lagrangian equations of motion
(B92, BE93, B94, EB97) involves taking the Eulerian 
divergence and curl of the first equation, combining with the second and using the homogeneous equations for
the scale factor $a(t)$ to give
\bea
\nabla_{r} \cdot {\ddot {\bf  r}} &=& -4 \pi  G \left[ \rhom + \rhoXav (1+3 w) \right] \label{diveq}\\
\nabla_{r} \times {\ddot {\bf  r}}&=&0, \label{curleq}
\eea
where $\rhom$ and $\rhoXav$ are the matter and dark energy
densities respectively.  These steps can eliminate 
the potential and operator occurrences of ${\bf
  x}$ or ${\bf r}$ on the right-hand side.  The essential
complication, however, is that the derivative form of the 
geodesic equations is insensitive to a spatially homogeneous
arbitrary, time-dependent vector shift $\Delta {\bf r}(t)$.

The solution of the resultant Lagrangian system by Ehlers
and Buchert (EB97) is achieved in a specific frame, one in which
the initial, spatially averaged peculiar velocity vanishes.
We must address the relationship between the solutions of
the geodesic equation and of this specific solution of the
derivative form of the geodesic equation.  To handle this
situation carefully, we start in the observer frame
(denoted by OBS) and introduce the Ehlers-Buchert frame (denoted by
EB). The solution to the geodesic equation lies in the
former and the specific solution to the derivative form in
the latter. The connection is called the frame shift.

Schematically, the initial data is specified in the OBS
frame, transformed to the EB frame, solved in the EB frame
using EB's perturbative expansion and then transformed back
to the OBS frame. The EB frame is not an inertial frame and the transformation between the two frames is not a Galilean transformation. We
regard the EB frame as a computational frame; it coincides with the physical observer frame only at the initial time. 

\subsection{Notation for Observer and EB Frames}

Begin by assuming that each frame possesses its own comoving
coordinate system. Imagine
a single particle variously described according to multiple
coordinate systems.  In each
frame the Lagrangian label is assigned to be the comoving
Eulerian coordinate at the initial time
\bea
{\bf Y} &=& {\bf x}_{OBS,0}\\
{\bf X} &=& {\bf x}_{EB,0}.
\label{defn}
\eea 
At future times the particles positions are written
${\bf x}_{OBS}({\bf Y},t)$ and ${\bf x}_{EB}({\bf X},t)$, respectively.
Physical coordinates, Jacobians relating Eulerian to Lagrangian coordinates
and volume elements are defined in the usual manner in the OBS frame
\bea
{\bf r}_{OBS}({\bf Y},t) &=& a(t) {\bf x}_{OBS}({\bf Y},t) \\
J({\bf Y}, t) & = & 
{\rm Det} \left (\frac{\partial {\bf r}_{OBS}}{\partial {\bf Y}}\right ) \\
d^3x_{OBS} & = & \frac{J({\bf Y},t)}{a^3} d^3Y 
\label{eq:jacobtrans}
\eea
and it follows $J({\bf Y}, t_0) = \ainit^3$. Similar definitions
are given in the EB frame.

Without loss of generality
choose the two frames to coincide at the initial time,
${\bf x}_{OBS}={\bf x}_{EB}$ and ${\bf Y} = {\bf X}$, and write
the relationship between the comoving coordinates as
a pure function of time
\bea
\label{eq:posshift}
{\bf x}_{OBS}({\bf Y},t) &=& {\bf x}_{EB}({\bf X},t) + \Delta {\bf x}(t) .
\eea
Here, $\Delta {\bf x}(t)$ is the comoving frame shift.
Next, define $\Delta{\bf r}$ in terms of $\Delta{\bf x}$
\beq
\Delta{\bf r}(t)        = a(t) \Delta{\bf x}(t) .
\eeq
We now have
\bea
\Delta{\bf x}(t_0)      &=& \Delta {\bf r}(t_0) = 0 \\
J({\bf Y}, t) & =  & J({\bf X},t) .
\eea
Initially, $d^3 x_{OBS} = d^3 Y = d^3 X = d^3 x_{EB}$, and at all times
the Eulerian volume elements satisfy
$d^3 x_{OBS}=d^3 x_{EB}$ and the Lagrangian volume elements
$d^3 Y = d^3 X$.  We will choose ${\dot \Delta}{\bf x}(t_0)$ in a
moment and then later make use of the remaining freedom to choose $\Delta{\bf x}(t)$
so that the geodesic equation is solved.

The physical mass density is $\rho_m({\bf r},t)$. Since it's a scalar
it remains unchanged by the EB to OBS transformation.
Mass conservation and \eqnrefs{eq:rhoy} and \eqnrefbare{eq:jacobtrans} give
\beq 
\rho_m({\bf x}, t) a^3 d^3 x_{OBS} = \rho_m({\bf Y}, t_0) a_0^3 d^3 Y .
\label{eq:masscons}
\eeq
The peculiar velocity is defined
\bea
\label{eq:Epecvel}
{\bf v}_{OBS}({\bf Y},t) & = & {\dot {\bf r}}_{OBS}({\bf Y},t)- \frac{\dot a}{a} {\bf r}_{OBS}({\bf Y},t) \\
& = & a {\dot {\bf x}}_{OBS}({\bf Y},t) 
\eea
and likewise for ${\bf v}_{EB}$. The peculiar velocity shift is defined
\bea
\label{eq:velshift}
\Delta {\bf v}(t) & = & {\bf v}_{OBS}({\bf Y},t) - {\bf v}_{EB}({\bf X},t) \\
\label{eq:velshift2}               & = & {\dot \Delta} {\bf r}(t)  - \frac{\dot a}{a} \Delta {\bf r}(t) \\
\label{eq:velshift3}         & = &  a {\dot \Delta {\bf x}}(t) .
\eea

\subsection{Transformation from Observer to EB Frame}
\label{subsec:OBStoEB}
The perturbation is characterized by two quantities specified in the observer frame; the initial fractional overdensity 
\beq
\delta_{OBS}({\bf Y}, t_0)= \frac{\rhomofYtzero}{\rhominit} -1
\label{eq:deltadef}
\eeq  
and the initial peculiar velocity 
\beq {\bf v}_{OBS}({\bf Y}, t_0) = {\dot {\bf r}}_{OBS}({\bf Y},t_0) - {\dot a}_0 {\bf Y}. 
\label{eq:vpec}
\eeq 
The density perturbation is a scalar quantity so transferring to the EB
frame is immediate
\beq
\delta_{EB}({\bf X},t_0) = \delta_{OBS}({\bf Y},t_0) 
\label{deltaEBdef}
\eeq
and we will henceforth drop the ``subscript'' that distinguishes these
two.

The initial peculiar velocity is a vector quantity which transforms like
\beq
{\bf v}_{EB}({\bf X},t_0) = {\bf v}_{OBS}({\bf Y},t_0) - {\bf v}_{c,0}
\label{velEB}
\eeq
where ${\bf v}_{c,0}$ is a constant we will now choose. Let
$\langle \cdots \rangle_{s} \equiv (\int d^3s ...)/\int d^3s$ denote the average of a quantity over the variable `$s$'. In this notation, the average peculiar velocity in the OBS frame at {\it any time} $t$ is 
\beq
{\bf v}_{c}(t) = \langle {\bf v}_{OBS}({\bf x}_{OBS},t) \rangle_{x_{OBS}}. 
\eeq
We want to ensure that the average peculiar velocity in the EB frame vanishes at the
beginning of the step,
$\langle {\bf v}_{EB} \rangle_{X} = 0$, as assumed in the
formal perturbative scheme given in EB97 (see \S \ref{subsec:scheme}). 

From \eqnref{velEB}, this requirement sets $\Delta {\bf v}(t_0)={\bf v}_{c,0} = {\bf v}_c(t_0)$. At the beginning of the step the Eulerian and Lagrangian coordinate systems coincide,
$\Delta {\bf r}(t_0)=0$, 
and the initial time derivative of the frame shift is
\beq
{\dot \Delta}{\bf r}(t_0) = 
\ainit {\dot \Delta}{\bf x}(t_0) = {\bf v}_{c,0}.
\label{eq:initfrshiftvel}
\eeq

\capeqnref{velEB} refers to the initial condition whereas \eqnref{eq:velshift} applies at all times during a step. In \S \ref{frameshifts} we give explicit expression for the shift accumulated during a time $t$. If one takes a single step then ${\bf v}_{c,0}$ is computed once at the start of the evolution. At the end of the step accumulated position and velocity shifts are added to the EB frame results to obtain the solution in the OBS frame. If one re-expands the solution, one takes multiple steps and repeats the process over and over.

\subsection{Equations in EB Frame}

Conservation of mass and \eqnrefs{defn}, \eqnrefbare{eq:deltadef} 
and \eqnrefbare{deltaEBdef} imply
\beq
\delta({\bf X},t) = \frac{(1+\delta({\bf X},t_0)) a^3}{J({\bf X},t)} - 1.
\label{eq:delta}
\eeq
The density evolution of dark energy 
(denoted by a subscript $\darkenergysubscript$ 
not to be confused with the Lagrangian coordinate) for a constant equation of state $w$ is
\beq
\rhoXav  = \rhoXinit\left(\frac{\ainit}{a}\right)^{3(1+w)}.
\label{eq:deden}
\eeq
These allow one to express the right-hand side of \eqnref{diveq} in
terms of Lagrangian coordinates. We
transform all the derivatives with respect
to the Eulerian coordinates ${\bf r} \equiv {\bf r}_{EB}$ 
in \eqnref{diveq} and \eqnrefbare{curleq} to
derivatives with respect to the Lagrangian coordinates
(Appendix \ref{mathtrans} furnishes additional details). The
complete set of equations is
\bea
{\hat L}[{\ddot {\bf r}_{EB}},{\bf r}_{EB},{\bf r}_{EB}] &=&
-3\Hinit^2 \Omega_{\mattersubscript,0}\ainit^3(1+\delta({\bf X},t_0))
\label{mainL} \\
\nonumber && -\frac{\Hinit^2}{2}  (1+3 w) \Omega_{\darkenergysubscript,0} \left(\frac{\ainit}{a}\right)^{3(1+w)}{\hat L}[{\bf r}_{EB},{\bf r}_{EB},{\bf r}_{EB}], \\
{\hat {\bf T}}[{\ddot {\bf r}_{EB}},{\bf r}_{EB}] &=& 0,
\label{mainT}
\eea
where
\bea
\Omega_{\mattersubscript,0} &=&  \frac{8 \pi G \rhominit} {3 \Hinit^2},  \\
\Omega_{\darkenergysubscript,0}&=&   \frac{8 \pi G \rhoXinit} {3 \Hinit^2}
\eea
and the action of ${\hat L}$ and ${\hat {\bf T}}$ operators on general vectors ${\bf A}, {\bf B}, {\bf C}$ is defined as 
\bea
{\hat L}[{\bf A},{\bf B},{\bf C}] &=&  \epsilon_{lmq}\epsilon_{ijk} \frac{\partial A_i}{\partial X_l}\frac{\partial B_j}{\partial X_m}\frac{\partial C_k}{\partial X_q},
\label{Ldef}\\
{\hat T}_q [{\bf A},{\bf B}] &=& \epsilon_{lmq}\frac{\partial A_k}{\partial X_l}\frac{\partial B_k}{\partial X_m}.
\label{Tdef}
\eea
Here $\epsilon_{ijk}$ is the usual Levi-Civita symbol and
Einstein's summation convention is used. The scalar operator ${\hat L}$
and vector operator ${\bf \hat T}$ provide a compact
representation. 

The specification of $\ainit$, $\Hinit$,
$\Omega_{\mattersubscript,0}$,
$\Omega_{\darkenergysubscript,0}$, $w$, the initial
fractional overdensity and the initial peculiar velocity
determines the solution. The fractional overdensity appears
explicitly in the equations above; the peculiar velocity
enters as initial conditions for ${\dot {\bf r}_{EB}}$. The system defined by \eqnrefs{mainL} and \eqnrefbare{mainT} can be found in EB97 (see references therein for prior versions) for the case of a cosmological constant. The generalization here allows for time-dependent dark energy term.

\subsection{EB Perturbation scheme}
\label{subsec:scheme}
In the formalism outlined by B92, BE93, B94 and EB97, the physical coordinate is split
\beq
{\bf r}_{EB}({\bf X},t)= a(t) {\bf X} + {\bf p}({\bf X},t)
\eeq
where the first term is the position in a homogeneous
universe (zero density perturbations and zero peculiar velocities)
and the second represents displacements that occur in
the general case. The choice of the initial
comoving coordinate as the Lagrangian label (equation \ref{defn}) implies ${\bf p}({\bf  X}, t_0) = 0$. 

The displacement vector is expanded 
\beq
\label{fexpansion}
{\bf r}_{EB}({\bf X},t)= a(t) {\bf X} + \sum_{n=1}{\bf p}^{(n)}({\bf X}, t) \epsilon^{n},
\eeq
where $\epsilon$ is the formal, bookkeeping device that tracks the
order of the displacement. Substitute the ansatz \eqnref{fexpansion} into
\eqnrefs{mainL} and \eqnrefbare{mainT} and equate the terms
of the same order of $\epsilon$ to generate a hierarchy of
equations to be solved. Since the Lagrangian labelling is
independent of time the derivatives of interest commute:
$\left[ d/dt, \nabla_X \right] = 0$. This makes it
straightforward to find ${\dot {\bf r}}_{EB}$ and ${\ddot {\bf r}}_{EB}$.

This formal treatment at {\it zeroth order} of \eqnref{mainL} reduces to
\beq
\frac{ {\ddot a}}{a} = -\frac{\Hinit^2}{2} \left(\frac{\Omega_{\mattersubscript,0} \ainit^3}{a^3} + (1+3 w) \Omega_{\darkenergysubscript,0} \left(\frac{\ainit}{a}\right)^{3(1+w)} \right) 
\label{zerothorder}
\eeq
while \eqnref{mainT} is identically zero. This is simply the
equation governing the background scale factor. Given
$\ainit$, $\Hinit$, $\Omega_{\mattersubscript,0}$,
$\Omega_{\darkenergysubscript,0}$ and $w$ the background evolution
is completely determined.

At first order \eqnrefs{mainL} and \eqnrefbare{mainT} reduce to
\bea
D_t^L \left[ \nabla_{\rm X}\cdot{\bf p}^{(1)} \right] &=&-\frac{3}{2}\Hinit^2\Omega_{\mattersubscript,0}\ainit^3\delta({\bf X},t_0)
\label{firstorderL}\\
D_t^T \left[\nabla_{\rm X} \times {\bf p}^{(1)}\right]&=&0
\label{firstorderT}
\eea
and at higher orders to 
\bea
\label{higherorderL}
D_t^L\left[ \nabla_{\rm X}\cdot{\bf p}^{(n)} \right] &=& S^{(n,L)} \\
D_t^T \left[ \nabla_{\rm X} \times {\bf p}^{(n)}\right] &= & {\bf S}^{(n,T)},
\label{higherorderT}
\eea 
where the operators are
\bea
\label{DtL}
 D_t^L &=& \left(2a \ddot a+\frac{3}{2}a^2 \Hinit^2 (1+3 w) \Omega_{\darkenergysubscript,0} \left(\frac{\ainit}{a}\right)^{3(1+w)} + a^2 \frac{d^2}{dt^2}\right),\\
\label{DtT}
D_t^T &= &\left(-\ddot a + a \frac{d^2}{dt^2}\right) 
\eea
and $S^{(n,L)}$ and ${\bf S}^{(n,T)}$ are scalar and vector source terms comprised of combinations of displacements with order $<n$. An explicit form is given in the Appendix \ref{app:initialcond}. Here L and T
refer to longitudinal and transverse operators.

Next, the displacement is split into
longitudinal and transverse parts, order-by-order:
\beq 
{\bf p}^{(n)} = {{\bf p}^{(n,L)}} + {{\bf p}^{(n,T)}}, 
\eeq
where $\nabla_{\rm X}\times {{\bf p}^{(n,L)}}=0$ and
$\nabla_{\rm X}\cdot {{\bf p}^{(n,T)}} =0$. On the 3-torus,
a unique decomposition for order $n$ requires
$\langle {\bf p}^{(n)}({\bf
  X},t) \rangle_{X} = 0$ (see Appendix C of EB97 for the mathematical proof).
The choice of Lagrangian labels guarantees the condition is
satisfied at $t=t_0$ for all $n$. If the volume-averaged source terms
vanish at all times then the decomposition is also possible
at future times. Appendix \ref{zeromean} shows that
$\langle {\bf p}^{(1)}({\bf  X},t) \rangle_{X} = 0$ 
at future times if both
the overdensity and the peculiar velocity average to zero {\it initially},
which is the case (the latter by the choice of the EB frame). 
Using the $n=1$ results
one next shows that the average scalar and vector source terms for $n=2$
displacements vanish so 
$\langle {\bf p}^{(2)}({\bf  X},t) \rangle_{X} = 0$ and this argument
is extended order-by-order. The
result is that the requirements for decomposition are satisfied
at all orders.

Separating longitudinal and transverse displacements order-by-order
uncouples the left-hand sides of 
\eqnrefs{higherorderL} and \eqnrefbare{higherorderT} and
rationalizes the ``L'' and ``T'' labelling {\it a posteriori}.
There are separate source term ($S^{(n,L)}$, ${\bf  S}^{(n,T)}$) but
note that each depends upon both types of lower order solutions 
(${\bf p}^{(m,L)}$ and ${\bf p}^{(m,T)}$ for $m<n$). The entire solution
at lower orders is needed to compute either the
longitudinal or transverse source for a given order.

The specification of the initial conditions for the
hierarchy of equations determines the physical meaning of
the formal perturbative expansion in powers of
$\epsilon$. We make specific choices to ensure
that the initial density and
velocity perturbations are first order in $\epsilon$. 
The first choice is ${\bf p}^{(n,L/T)}({\bf
  X},t_0) = 0$ for all $n$ (``L/T'' means the equation
applies to both longitudinal and
transverse displacements) which is sufficient but not
necessary to insure zero initial displacement. 
Homogeneous initial data makes it easy to use the formal structure
of the hierarchy to check how powers of density and velocity
enter.  The density appears explicitly only at first order, i.e.
in the equation for ${\bf p}^{(1,L)}$.  The second choice is
that the peculiar velocity is assigned as initial data only
to the derivative of the first order term
\bea
{\dot {\bf p}}^{(1,L/T)}({\bf X},t_0) &=& {\bf v}_{EB}^{L/T}({\bf X}, t_0)
\eea
where ${\bf v}_{EB}^{L/T}({\bf X}, t_0) $ are the curl-free
and divergence-less parts of the initial velocity. For all
$n>1$
\bea
{\dot {\bf p}}^{(n,L/T)}({\bf X},t_0) &=& 0 .
\eea
One might contemplate other choices (i.e., one could deviate
from homogeneous for the displacements and/or distribute the
peculiar velocity to different orders) but these would not
have the simple physical interpretation that the natural
choices provide and could vitiate the split into
longitudinal and transverse components.

In \eqnrefs{firstorderL}, \eqnrefbare{firstorderT},
\eqnrefbare{higherorderL} and \eqnrefbare{higherorderT} the
spatial and temporal operators commute and the
spatial and temporal parts of the solution decouple at each order (see
Appendix \ref{app:initialcond}). The solution is a sum
of spatial parts times temporal parts. Each spatial part involves
solving linear, elliptic partial differential equations of
the form $\nabla \cdot {\bf F} = S^L$ or $\nabla \times {\bf
  F} = {\bf S}^T$ with known sources $S^L$, ${\bf S}^T$ (Poisson-like
problems). Each temporal
part involves solving a second order ordinary differential
equation in time (initial value problems). Both sets of equations
include coefficients that depend upon the background
cosmology.

As a practical matter all the Poisson-like equations are
solved using Fourier transforms on a $N\times N \times N$
grid with equally spaced grid points which represent the
Lagrangian coordinates. Initial value equations
are solved numerically using a standard differential
equation solver. The individual displacement terms are
summed to reconstruct ${\bf r}_{EB}({\bf X},t)$ using
\eqnref{fexpansion}.

\subsection{Frame shifts}
\label{frameshifts}

The LPT expansion described in the previous section yields
the position, velocity and density in the EB frame. To
get the corresponding answers in the observer frame one must
add the time dependent frame shifts (equations \ref{eq:posshift}
and \ref{eq:velshift}). 
 
Start with the Newtonian limit of the geodesic equation in the observer frame 
\beq 
\frac{d}{dt}\left(a^2 \frac{d {\bf x}_{OBS}}{dt} \right) = -{\bf \nabla}_{x_{OBS}} \psi ({\bf x}_{OBS},t)
\label{geo2}
\eeq
and take the average with respect to the Eulerian ${\bf x}_{OBS}$
(on the left we would formally replace
$d{\bf x}_{OBS}/dt$ with the local Eulerian velocity and
$d/dt$ with the full convective derivative).
The right-hand side is periodic and averages to zero. Now formally we
return to the Lagrangian description on the left:
substitute for ${\bf x}_{OBS}$ from \eqnref{eq:posshift} and use 
$d^3 x_{OBS} = d^3 x_{EB} =  J({\bf X},t) a^{-3} d^3X$  to give
\beq
\frac{d}{dt}\left ( a^2 \frac{d \Delta {\bf x}(t)}{dt} \right) =   - \frac{1}{V} \int \left[ \frac{d}{dt} \left( a^2 \frac{d {\bf x}_{EB}}{dt} \right) \right]  a^{-3} J({\bf X},t)    d^3X.
\label{eqforshift}
\eeq
and $V = \int d^3 x$ is the comoving volume of the box (physical volume
$a(t)^3 V$).

The geodesic equation is satisfied if the shift $\Delta {\bf x}$ satisfies
this second order ODE with initial conditions $\Delta {\bf x}(t_0) = 0$ and ${{\dot \Delta {\bf x}}}(t_0) = {\bf v}_{c,0}/\ainit$. The solution is 
\beq
\Delta {\bf x}(t) = \int_{t_0}^t \left( \frac{1}{a^2}\int_{t_0}^{t'} {\bf g}(t'') dt'' \right) dt' + \int_{t_0}^t \frac{{\bf v}_{c,0} \ainit}{a^2} dt' 
\label{shiftsoln}
\eeq
where 
\bea
{\bf g} (t) &=& - \frac{1}{V} \int \frac{d}{dt} \left(a^2 {\dot {\bf x}}_{EB} \right) (a^{-3}J) d^3X \\
&=&  - \frac{1}{V} \int (a {\ddot {\bf p}} - {\ddot a}   {\bf p}) (a^{-3} J) d^3 X.
\eea
Note that
${\bf g}(t)$ is completely determined from the EB solution.

In the observer's frame the physical position and velocity are 
\bea
{\bf r}_{OBS}({\bf Y}, t) &=&a {\bf X} + {\bf p}({\bf X},t)  + a \Delta {\bf x}(t) 
\label{robs}\\
{\dot {\bf r}}_{OBS}({\bf Y}, t) &=&{\dot a} {\bf X} + {\dot {\bf p}}({\bf X},t)  + {\dot a} \Delta {\bf x}(t) + a {\dot \Delta {\bf x}} (t)
\label{rdotobs}
\eea
and hence the peculiar velocity (defined by \eqnref{eq:Epecvel}) is 
\beq
{\bf v}_{OBS}({\bf Y},t) = {\dot {\bf p}}({\bf X},t) -\frac{{\dot a}}{a} {\bf p}({\bf X},t)  + \frac{{\bf v}_{c,0}\ainit}{a} +  \frac{1}{a}\int_{t_0}^{t} {\bf g}(t') dt'.
\label{vobsofY}
\eeq
The density in the observer frame is identical to that in the EB frame
\beq 
\delta_{OBS}({\bf Y},t) = \delta_{EB}({\bf X},t) 
\eeq
where the right-hand side is given by \eqnref{eq:delta}. 

Note that the frame shift is time-dependent. The right hand side of \eqnref{shiftsoln} has two integrals: the first arises from requiring that the geodesic equation is satisfied whereas the second is the decay (due to Hubble drag) of the non-zero average initial velocity defined in \S \ref{subsec:OBStoEB}. The shift is not a fixed Galilean transformation. It is related to the fact that the system given by \eqnrefs{diveq} and \eqnrefbare{curleq} is `translation covariant' (Heckmann-Sch\"{u}cking transformation; see \citealt{buchert_class_1987} and references therein). In particular, the position, velocity {\it and} acceleration are different in the two frames. The density in the two frames, as inferred from the divergence of the acceleration, is the same. That operator is insensitive to any time dependent spatially uniform acceleration. Such accelerations are absent in the inertial frame but present in the computational one. Existing treatments of LPT set the frame shift to zero and satisfy \eqnrefs{diveq} and \eqnrefbare{curleq}. However, this does not yield the solution in the OBS frame. We will demonstrate in later sections that ignoring the frame shift destroys momentum conservation and convergence. 

\subsection{Frame shifts and momentum conservation}
\label{sec:Pcons} 

Frame shifts ensure that the geodesic
equation is properly satisfied. This is not just a
mathematical requirement but a physical one. In this section
we consider the implications for momentum conservation
and drop the explicit OBS subscript for clarity.

The total proper momentum in comoving coordinates written as
a sum over mass elements $\Delta M_i$ or an integral over
Lagrangian labels (i.e. the density at the initial time) is
\bea
{\bf P}(t) & = & \sum_i a \frac{d {\bf x}_i}{dt} \Delta M_i \\
           & = & \int a \frac{d {\bf x}}{dt} \rho_m({\bf Y},t_0)a_0^3 d^3Y .
\label{Pdefn}
\eea
where $x = x({\bf Y},t)$. Multiply by $a$ and take the time derivative
\bea
\frac{d (a {\bf P})}{dt} & = & \int {\bf Q} \rho_m({\bf Y}, t_0) a_0^3 d^3Y , \\
{\bf Q} & = & \frac{d}{dt}\left(a^2  \frac{d {\bf x}}{dt}\right).
\eea 
Now replace the variable of integration ${\bf Y}$ with
$x=x({\bf Y},t)$ at the time of interest. This recasts the
integration from a Lagrangian to an Eulerian form. Using
\eqnref{eq:masscons} the densities are $\rho_m({\bf Y},t_0)
a_0^3 d^3Y = \rho_m({\bf x}, t) a^3 d^3 x = (\rho_m(t) +
\delta\rho_m({\bf x}, t)) a^3d^3x$ and in the last equality
we have introduced the background and the fluctuating
densities. The integral splits
\bea
\frac{d (a {\bf P})}{dt} & =  &
\rho_m(t) a^3 \int {\bf Q} d^3x   + a^3 
\int {\bf Q} \delta\rho_{m}({\bf x},t)  d^3x.
\eea
When the geodesic equation \eqnref{originaleq} is satisfied
${\bf Q}$ is proportional to $\nabla
\psi$. Since $\psi$ is periodic the first volume integral
with integrand proportional to ${\bf Q}$ vanishes. When
the Poisson equation \eqnref{originaleqII} is satisfied one
may rewrite $\nabla \psi$ using the formal Green's function
for $\psi$ and invoke its symmetries to show that the second
integral vanishes. Hence, $d (a {\bf P})/dt = 0$ and ${\bf
  P} \propto 1/a$.

If Poisson's equation is solved exactly in the EB frame the
second integral will vanish. However, if frame shifts
are ignored then the geodesic equation will not be satisfied
and the first integral will generally fail to vanish. 

A qualitative understanding of why a non-zero ${\bf v}_c$
must develop is revealed by looking at a problem with
Zel'dovich initial conditions. We characterize this system
as starting with first order terms in displacement,
peculiar velocity and peculiar acceleration strictly proportional to
one and other and negligible higher order terms
(see Appendix \ref{App:zeldovich} for a detailed discussion).
Using ${\bf v} = a {\dot {\bf
    x}}$, $\rho_m(t) = \rho_{m,0} a_0^3/a^3$ and the
definition of $\delta$, the total momentum is
\beq
{\bf P}(t)  =  \rho_{m,0} a_0^3 V \left({\bf v}_c(t)   + \frac{1}{V} \int {\bf v} \delta({\bf x},t) d^3x 
\right)
\eeq
where $V$ is the comoving volume of the box.  The initial state
with velocity proportional to acceleration implies that the first term
${\bf v}_c=0$ because the system is periodic and the mean acceleration vanishes. 
The second term $\propto \int {\nabla \psi}
\delta d^3x = 0$ on account of the symmetries of the Green's
function for Poisson's equation. Hence, Zel'dovich initial
conditions in a periodic system imply ${\bf
  P}(t_0)=0$. If/when the system evolves so that ${\bf v}$
is no longer proportional to acceleration then the second
term will no longer vanish. The first term must be present
and non-zero to insure ${\bf P}(t)=0$. In summary, the EB
frame's non-zero velocity is intimately tied to the momentum
conservation law which is maintained by proper inclusion of the frame shifts. 

Whenever a single step, first order scheme is used to initialize
an N-body calculation starting with Zel'dovich initial
conditions the frame shift vanishes, i.e. $g(t)=0$, accurate at first order.
To see this insert the definition of $J$ from \eqnref{jacobdef},
the proportionality of displacement, velocity and acceleration
into \eqnref{shiftsoln}.  In general, a
second order scheme that omits the frame shift completely
makes a second order mistake, i.e. is accurate at first
order but not second. 

This example also helps one understand some special
situations when frame shifts are exactly zero. Assume the initial
density and velocity field are purely one dimensional i.e.,
parallel planes of matter in 3D. As long as particle
trajectories do not cross the acceleration on a particle is
fixed by the initial density field. A single step first
order LPT scheme yields an exact description and it is
possible to write out explicit expressions for the frame
shifts. For this one dimensional limit the mean velocity (in
the OBS comoving coordinate system) is conserved i.e.,
$\langle {\bf v}\rangle_{x_{OBS}} \sim {\bf v}_{c,0} a_0 /a$. If it is
initially zero, it is always zero. This does {\it not} imply
that frame shifts vanish because there are two
contributions, one coming from ${\bf v}_{c,0}$ and the other
from internal dynamics in \eqnref{shiftsoln}. 
In one dimension both terms will vanish if the
initial velocity is proportional to the initial
acceleration. In this special situation the observer and EB
frames coincide at all times and the frame shift terms are
exactly zero for both single and multi-step schemes.  Proofs
are provided in Appendix \ref{App:1D}.  

Frame shifts are needed in other contexts and omitting
them will generally produce inconsistencies that manifest as
lack of momentum conservation.  To the best of our knowledge
these frame shifts have been ignored in most applications of
LPT so far.

Let's briefly consider the physical nature of ${\bf v}_c$.
If the box contains a fair sample of the universe and if the
comoving coordinate system coincides with the preferred FRW
frame (one with an isotropic cosmic microwave background
radiation) then one anticipates ${\bf v}_{c,0} = 0$. On the
other hand, if super-horizon motions are present (so the
notion of a fair representation is not satisfied) and/or if
the observer's frame is not equivalent to the preferred FRW
frame then, generally speaking, ${\bf v}_{c,0}$ is non-zero. In such cases 
the EB frame coincides with the OBS frame only at the beginning of the calculation. 

In an inhomogeneous cosmology ${\bf v}_{c}(t)$ is
not proportional to the mean momentum in the box.  Of
course, the box's self-generated gravitational forces cannot
impart net momentum to the box as a whole. If the initial
net momentum is zero it will always be zero, however, this
does {\it not} imply that ${\bf v}_c$ will vanish on future
Eulerian comoving grids if it vanishes initially. In
particular, if one chooses ${\bf v}_c(t_0)=0$ then at later
time $t_1$ one generally has ${\bf v}_c(t_1) \ne 0$.

As a practical matter if one employs the EB method of solution
one cannot circumvent the need to
start each step by finding the frame where ${\bf v}_{c}$
vanishes.  There is
no single, adroit selection to be made at the calculation's
start because the EB frame is not tied to a conserved
quantity.

\subsection{LPT re-expansions}
\label{re-expansions}

The previous two sections describe the complete formalism
for single step LPT; start with initial densities and
velocities in the OBS frame, transform to the EB frame,
solve for the displacements, move back to OBS frame and
compute the densities and velocities at the end of the step.
Two questions remain. How big a step can be made and what to
do if that step doesn't encompass the total evolutionary
time of interest?

In NC11 we answered those questions in the context of
spherically symmetric perturbations in an $\Omega=1$
cosmology. We demonstrated that all convergence problems
associated with the LPT series could be overcome by
re-expanding the series in overlapping time domains with
each domain subject to a time of validity criteria. The time
of validity was determined rigorously as functions of the
initial density and velocity perturbations.

Here, we assume that the time of validity for the
inhomogeneous evolution can be estimated by treating each
point in the box as if it were an isolated top-hat. That is,
we use the local density perturbation and local velocity
perturbation to calculate the time interval that would be
allowed for a top-hat. We find the maximum time that lies
within all the individual intervals throughout the volume
and set the time step accordingly.

The fractional overdensity $\delta$ and the fractional
Hubble parameter $\delta_v$ are the specific parameters
used to characterize the time of validity for the
spherical perturbation. For generic inhomogeneous initial
conditions we write the natural generalizations
\bea \delta &\equiv& \delta({\bf Y},t_0) 
\label{eq:deltadefn} \\
\delta_v& \equiv & \frac{1}{3 \Hinit} \nabla_{{\bf r}} \cdot {\dot {\bf r}({\bf Y} ,t_0)} -1\\
& = & \frac{1}{3 \dot \ainit} \nabla_Y \cdot {\bf v}({\bf Y},t_0) .
\eea
The values for $\delta$ and $\delta_v$ are identical in OBS and EB frame.
The time of validity $T(\delta, \delta_v)$ was determined in NC11;
we use those results to find the minimum of $T$ over the Lagrangian grid 
\footnote{The spherical top-hat system has no transverse component so
our treatment ignores that additional complication present 
in the inhomogeneous system. We see no evidence that this omission
leads to any practical difficulty in stability or in convergence.}.

A positive $\delta$ at a point implies an overdense region and a
positive $\delta_v$ implies an expanding region. From our experience
with top-hat evolution we know that if the time
of validity is set by an expanding region then the LPT
re-expansion scheme has the ability to 
extend the solution arbitrarily far into
the future. If it is set by
a collapsing region then the LPT re-expansion can extend the
solution to caustic formation. We remind the reader that the 
re-expansion scheme does not include
any physics for treating hot (multi-stream) fluid.

In LPT re-expansion the quantities calculated at the end of one
step form initial conditions for the next step. Since
each step begins with the Lagrangian coordinates equal to
the comoving coordinates there is
one additional computation that is necessary. A new Lagrangian
coordinate ${\bf Y}_1$ must be defined at the start of the next step;
this is related to the Lagrangian coordinate of the previous
step ${\bf Y}_0$ by
\beq
{\bf Y}_1 = \frac{{\bf r}_{OBS}({\bf Y_0},t_1)}{a(t_1)}, 
\label{Y0toY1}
\eeq
where $t_1$ is the starting time for the step and ${\bf
  r}_{OBS}({\bf Y_0},t_1)$ is given by \eqnref{robs}.  The
final quantities from the previous step are known on a
uniform grid in ${\bf Y}_0$; they are
transformed so that the initial conditions are specified
on a uniform grid in ${\bf Y}_1$. We
solve \eqnref{Y0toY1} for ${\bf Y}_0$ given
uniform ${\bf Y}_1$ and interpolate the final
quantities at ${\bf Y}_1$. We refer to this
step as regridding.

As explained in \S\ref{subsec:OBStoEB}, ${\bf v}_{c,0}$ differs for each step. Utilizing a sequence of frame shifts does not introduce any intrinsic errors. If the calculation were done exactly using infinite order LPT then the frame shifts at the end of a step would be computed exactly and, in turn, would yield the exact solution in the OBS frame. 
The re-expansion step involves interpolating the final density and velocity in the OBS frame onto a uniform grid. If interpolation were done without error then 
the new ${\bf v}_{c,0}$ at the start of the step would also be exact. In finite accuracy calculations ${\bf v}_{c,0}$ is contaminated by errors that are completely analogous to those in density or velocity. The entire solution converges when the tolerances are tightened.

\section{Numerical tests of the code}

\label{sec:tests}
This section checks and verifies the theoretical scheme
outlined in \S\ref{sec:setup} by carrying out selective
numerical calculations. 
The truncation error in the calculations depends upon the
Lagrangian order ($n$), the number of time steps ($N_t$),
and the size of the spatial grid ($N_s$). We will also refer
to $N_t$ as the re-expansion frequency.

The basic elements of the calculation are:
\begin{enumerate}
\item Use FFT (Fast Fourier Transform) methods to solve all spatial equations,
\item Use Runge-kutta integration to solve the ordinary
  differential equations for the time-dependent
  coefficients (including the cosmology background),
\item Use Simpson's rule for the quadratures needed to evaluate
  the frame shifts (eq. \eqnrefbare{shiftsoln}),
\item Use Newton's method for root finding and exact
  periodic interpolation during the regridding
  step (the most time-consuming task).
\end{enumerate}
Generally, we complete each individual task to machine
precision. In the following section ``interpolation error''
refers to the net total of these four sources of numerical
error. The total is typically dominated by the roundoff
error in the solution of the spatial equations. These
manipulate matrices with $\sim N_s^3$ elements.  We observe
residual errors at roughly the level $\sim N_s^{3/2}
\epsilon_{Machine}$ where $\epsilon_{Machine}$ is the
machine precision. The interpolation error is small compared
to truncation error in all results we will discuss. The
truncation error is the main focus of investigation.

\subsection{Convergence and truncation error}

Consider evolution over a finite length of time prior to
particle crossing in a periodic box of comoving size $L$.
There are two qualitatively distinct LPT schemes that, in
principle, are capable of converging to the exact answer:
\begin{enumerate}
\item Fix $N_t$ and increase $n$ and $N_s$ without bounds.
\item Fix $n$ and increase $N_t$ and $N_s$ without bounds.
\end{enumerate}
In both schemes it is necessary to refine the
representation of all spatial functions
($N_s \to \infty$). In both it is necessary to
respect the time of validity for evolution. 
The first scheme is satisfactory
if $N_t$ is sufficiently large; the second scheme which
decreases the size of the time steps as $N_t \to \infty$ will
eventually respect {\it any} non-zero time of validity
condition.  Hybrid forms in which Lagrangian order and
number of steps increase together are also possible.

Achieving a theoretical understanding of the scaling of the
errors is important for two reasons. The numerical methods
should demonstrate the expected scaling.  Agreement between
expected and actual scaling is valuable information
when one is validating a computer code.  In addition, a
theoretical understanding helps make the most efficient
choices of control parameters for carrying out practical
calculations.

In the past, LPT with $N_t=1$ and fixed $n$
has been applied to practical cosmological problems while
the formal convergence issues have generally been
ignored. Occasionally, a solution has been studied as a
function of the Lagrangian order $n$ for a single step as in
the first scheme (\citealt{buchert_performance_1997}). There are
numerous comparisons with N-body calculations
(\citealt*{buchert_testing_1994},
\citealt*{melott_testing_1995},
\citealt*{karakatsanis_temporal_1997}) but these are not sufficiently
accurate to address the issue of LPT convergence. We will provide a
detailed look at convergence with respect to $n$ and $N_t$.

We treat $N_s$ in a somewhat different manner.  The spatial
representation within the box should be refined as order
and/or number of steps increase but we are unaware of any
study of LPT that did so.  Here, we will
study one example with smooth initial conditions to
establish that for sufficiently large grids the dominant
error is set by Lagrangian order and is
insensitive to grid size. For the remaining applications we
argue (on a case by case basis) that grid-related errors are
so small that convergence can be studied while holding $N_s$
large and fixed.

This strategy is a reasonable but not rigorous approach to
testing a method's convergence in a comoving,
periodic volume.  It is also important to note that making
the comoving box larger may be crucially important for
achieving accurate physical results. This is distinct from
solving a given problem in a given box. For
example, if the box holds a finite subset of a larger
physical system (e.g. the universe) and one is interested in
mean quantities of the larger system then one might need to
consider ever bigger boxes. Or, if the physical system in the
box obeys different boundary conditions
(e.g. isolated not periodic) then one might need to increase
the physical size. In these situations convergence to the
physical answer of interest will require that the grid density
$N_s/L$ and box size $L$ increase without bounds.

\subsection{Initial conditions}
\label{subsec:zeldovich}

At the initial time $t_0$, the system is completely
specified by the choice of cosmology, the initial density
field $\delta_{OBS}({\bf Y},t_0)$ and the peculiar velocity
${\bf v}_{OBS}({\bf Y}, t_0)$. Table \ref{paramtable} gives
a list of various parameters for the test runs presented
here. 
\begin{enumerate}
\item All tests are done in an $\Omega=1$ Einstein-deSitter
  cosmology over time intervals short enough that
  perturbations do not collapse to form multi-stream flows.
\item The initial conditions satisfy ${\bf v}_{c,0}
  =0$. This choice of convenience allows
  us to characterize the peculiar velocity initial
  conditions completely and directly in terms of transverse
  and longitudinal content. It has no essential impact on the
  general conclusions derived from the testing.
\item The tests cover both special and generic initial conditions.
  Special initial conditions possess non-trivial symmetries.
  For example, it is well-known that transverse modes decay
  with time in expanding cosmologies.  Practical
  cosmological simulations typically begin by setting ${\bf
    v}_{EB}^{T}({\bf X},t_0)=0$; such initial conditions
  are special by this criterion.  We also consider
  generic initial conditions to exercise the full dynamics.
\end{enumerate}

\begin{center}
\ra{1.3}
\begin{table}
\caption{Table of parameters for the numeric tests presented
  in \S\ref{sec:tests}. Here, 
  $n$ is Lagrangian order, 
  $N_t$ is number of geometrically spaced steps,
  $N_s$ is spatial grid size, 
  $\Delta t$ is the total time,
  $t_0$ is the initial time.
  The rms sub-scripted quantities are root mean square of
  $\delta$, the initial density contrast,
  ${\bf F}_\delta$, the initial first order acceleration, and
  ${\bf v}^{(L/T)}$, the initial velocity (longitudinal, transverse).
  Each test is discussed in the Section noted.
  1D means a one dimensional density and velocity profile.
  `Top-hat' means a compensated top-hat smoothed by a fixed
  Gaussian and including a non-radial, smooth velocity perturbation. `Exact' refers to the analytic solution given
for special initial conditions (Model I in \citealt{buchert_performance_1997}).}
\label{paramtable}
\begin{tabular}{ |c|c| c| c|c|c|c|c|c|c|c|} 
\hline
Section & Type & $n$ & $N_t$ & $N_s$ & $\frac{\Delta t}{t_0}$ & $\delta_{rms} $ & ${\bf F}^\delta_{rms}$ &${\bf v}^L_{rms}$ &${\bf v}^T_{rms}$\\
\hline
\ref{sec:1dexample}&1D&1 &  1-4 & 16 &  0.75 &0.007 & 0.112 & 0.007 & 0\\
\hline
\ref{sec:exactsoln}&Exact &3 &  1 & 16 &  .5 & 1.22 & 0.11 & 0.11 & 0\\
\hline
\ref{sec:gridNs}&Top-hat& 2 & 1 & 24-64 & 0.75  & 0.72 & 0.05 & 0.03  &0.042\\
\hline
\ref{sec:cosmocase}&Zel'dovich&1-3 &1-5 & 16&0.75  & 0.077 & 0.687 & 0.687&0\\
\hline
\ref{sec:gencase}&Generic& 1-3&1-6 &16  & 0.75&  0.077& 0.687  &0.270 &0.523\\
\hline
\end{tabular}
\end{table}
\end{center}

\subsection{Evolution of 1D perturbations}
\label{sec:1dexample}

First order LPT is well-known to be an exact method for
treating one dimensional problems prior to particle
crossing (\citealt{peacock99}). An ``exact'' solution is directly found by
quadrature irrespective of the identification of the EB
frame.  Here we compare exact results of this sort to
answers generated by our general 3D numerical implementation
of LPT re-expansion.

This comparison serves as a minimal check on the
full code at first Lagrangian order because it involves all
the basic elements of the most general calculation: the
Fourier representation of spatial functions, solution of the
ordinary differential equations fo`r the time-dependent
coefficients in the LPT expansion, evaluation of frame
shifts and the interpolation from one Lagrangian grid to
another. The solution of this simple problem exercises the
entire re-expansion LPT methodology. No special
modifications were made to the 3D code for carrying out this
test.

The initial density and velocity perturbations were taken to
be one dimensional functions consisting of a single Fourier
component $\delta({\bf X}, t_0) = 0.01 \cos 2 \pi X_1/L$ and
${\bf v}({\bf X}, t_0) = 0.01\{ \cos 2 \pi X_1/L,0,0\}$,
where $L$ is the comoving length of the box and $X_1$ is the
component along the $x$-axis. With this choice the initial
velocity and acceleration are {\it not} proportional and
we expect non-zero frame shift terms. Calculations
were made at Lagrangian orders $n=1-3$, for $N_t=1-4$ steps
covering a fixed total time $\Delta t$ with time steps $t_i
= t_0 (1 + \Delta t/t_0)^{i/N_t}$ for $0 \le i \le N_t$.
Regridding was done by solving \eqnref{Y0toY1} after each
step.

\begin{figure}
\includegraphics[height=5cm]{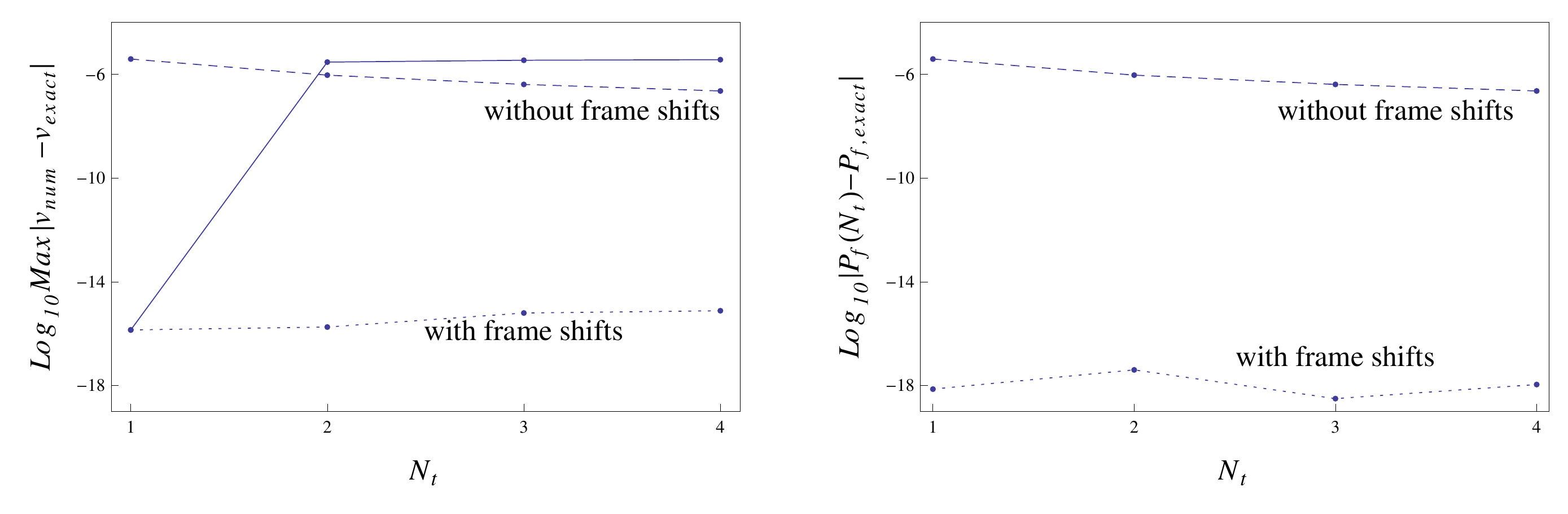}
\caption{Left hand panel: the dotted line shows the maximum
  difference of numerical and quadrature-derived, exact
  answers when frame shifts are included in the numerical
  answer; the dashed line shows the same information when
  frame shifts are omitted from the numerical answer.  Only
  when shifts are included do the numerical answers agree
  with the exact ones to machine precision. The solid line
  shows differences in single and multi-step numerical
  answers in the absence of frame shifts (none of these are
  ``exact''). Right hand panel: test of momentum
  conservation for the numerical runs with (dotted) and
  without (dashed) frame shifts.}
\label{onedexam}
\end{figure}

Exact and numerical results are compared in the observer's
frame at the final time.  As expected, the contributions to
the solution at second and third order were zero to machine
precision. The left panel of \figref{onedexam} shows the errors
(maximum absolute velocity difference on the grid
between exact and numerical results) as
a function of the numbers of steps taken to reach the final
time.  The lower dotted line at roughly the round off level
of computation (marked ``with frame shifts'') shows that
essentially perfect agreement with the exact answer is
achieved for both single and multiple steps. This result
validates (at first order) the implementation of LPT
re-expansion.  Note that a sequence of exact first order
calculations is exact.

Frame shifts play an essential part in the numerical
calculation. We repeated the calculation except that we imposed
$\Delta {\bf x}= \Delta {\dot {\bf x}} = 0$ in \eqnref{robs}
and \eqnref{rdotobs}. This omits the frame shifts. In left
panel of \figref{onedexam} the upper dashed line shows the
results obtained. There is an easily detected error with
respect to the exact answer, an error present even for
single step LPT.

Another useful comparison may be made between numerical single
and multi-step calculations even if one is ignorant of the
``exact'' solutions. As we have already shown above when
frame shifts are included single and multi-step first order
LPT calculations are equivalent. However, if one omits the frame
shifts this equivalency is broken.  The solid line shows the
difference between the numerical results for runs of a single step
and those of multiple steps when all frame shifts are omitted. The explanation
is simple: none of the answers is exactly right and the
discrepancies are tied to different numbers of missing frame shifts.

The right hand panel of \figref{onedexam} shows a test of
momentum conservation for calculations with and without frame
shifts and for varying numbers of steps.  With frame shifts
the numerical calculation achieves round off levels of
accuracy; without them the deficiency in the conservation
law is easily detected and, moreover, does not improve
as the interval in time is refined.

Taken together these simple tests help validate the method
and underline the crucial importance of including the frame
shifts.

\subsection{Exact solutions}
\label{sec:exactsoln}
An excellent check is Model I in \citet{buchert_performance_1997}.  It is an exact analytic
result for one step in the EB frame at the first three
orders for special initial conditions. Our numerical results
agree with the analytic results to machine precision for all
three orders. The special symmetry of the initial data
implies that many of the spatial functions of the full
general scheme vanish. The first, second and third orders
comprise 2, 3 and 16 non-zero terms, respectively, and
correspond to 2/3, 1/3 and 1/4 of the total number of
possible terms at each order. The agreement using the full
set of possible terms confirms that the both the time and
space pieces of non-zero terms are encoded correctly. The
agreement provides a consistency check for the spatial terms
that vanish but says nothing conclusive about the temporal
parts of these particular terms.

This check is essentially independent of the question of
frame shifts. The analytic expressions
for the frame shifts are exactly zero at all three orders. The numerically derived frame shift gave the same result to machine precision. It is not clear if the frame shifts continue to be zero at all higher orders, but for the purposes of this test we are concerned only with the first three orders. 

Ideally one would construct more general test cases to check
all possible terms in the code. Unfortunately, the method of
constructing `local forms' (B94)
is cumbersome and involves guessing the correct solution for
the Lagrangian displacement at each order so that the
divergence and curl match the source terms
order-by-order. We have utilized the checks against all the
published solutions of which we are aware.

\subsection{Top-hat with fixed smoothing}
\label{sec:gridNs}

The idealized configuration for this test is an overdense
sphere surrounded by a compensating region and an infinite
homogeneous background.  If the initial velocity
perturbation is zero the solution is simple and analytic.
There are two practical issues that potentially interfere
with running tests starting from this sort of initial data.
The computational volume is intrinsically periodic and the computational method 
is designed to handle continuous not discontinuous spatial
distributions.

A periodic configuration necessarily includes interactions
between neighbouring boxes because exact spherical symmetry
is not maintained in an approximate numerical solution. A
suitably compact configuration limits these errors and the
difference between the isolated and periodic versions can
be made negligibly small.
\footnote{The Fourier representation of the density
  guarantees that the mass within each box is exactly equal
  to the background value. The identical symmetries of the
  Cartesian grid within the box and of the arrangement of
  the surrounding periodic copies guarantee that the dipole
  interactions between copies vanish. Therefore, the leading
  interaction that distinguishes the isolated from the
  periodic problem is a quadrupole term. It is $\sim
  10^{-0.162 N_s } /L^4$ for the configuration studied. This
  is roughly equivalent to machine precision at $N_s =100$
  and is very small compared to the truncation errors
  studied here.} In principle, the top-hat is an excellent
problem for validating the LPT expansion since the
(isolated) solutions for the displacement and density can be
computed analytically {\it order by order}.  Numerical
results can be compared to analytic ones at a given order
(NC11). The significant complication is that the exact top-hat
profile is discontinuous along the boundaries between the
overdense sphere, the vacuum compensating region and the
outer region of average density. Such discontinuities induce
Gibbs phenomenon in the spatial Fourier representation and
mask differences between expected and observed solutions.
In principle, one can make the grid sufficiently large to overcome this
interference but we do not attempt to do so here because
the convergence is very slow. That means
we must rely on indirect checks that the higher order terms
in the LPT expansion are correct. 

Here, we smooth the top-hat density field by a Gaussian with
fixed width $\sigma$ and also impose a smooth velocity
field.  The resulting velocity profile has both longitudinal
and transverse velocities.  The details of the configuration
can be found in Appendix \ref{app:SPT}.  The
exact solution is unknown so we discuss convergence using
Cauchy differences for quantity $f$
\bea 
{\cal E}_f(\alpha;\alpha') \equiv
\sqrt{ \langle (f^{\alpha'} - f^{\alpha})^2 \rangle_{q} },
\label{cauchyerr1}
\eea
where $\alpha$ and $\alpha'$ label one or more of $n$, $N_t$
or $N_s$ (with the other parameters held fixed) and $q$ is
the coordinate system used.  We take $f=\delta$ for density
and $v$ for velocity at the final time. For $f=v$ all
three velocity components are included.

We calculate a single step of small size so that the final
time is well before collapse. The assessment of
convergence involves Cauchy differences formed with runs having
different grid resolutions and Lagrangian orders,
e.g. $\alpha=\{N_s,n\}$ and $\alpha'=\{N_s+1,n+1\}$. The
spectral representation of spatial functions suggests that if
grid errors dominate then the Cauchy differences will
decrease exponentially with $N_s$. Conversely, if errors in
the series expansion in LPT dominate then the scaling will be
exponential with $n$. This implies a sharp transition from
errors dominated by grid resolution to errors dominated by
Lagrangian order.

\begin{figure}
\includegraphics[width=6cm]{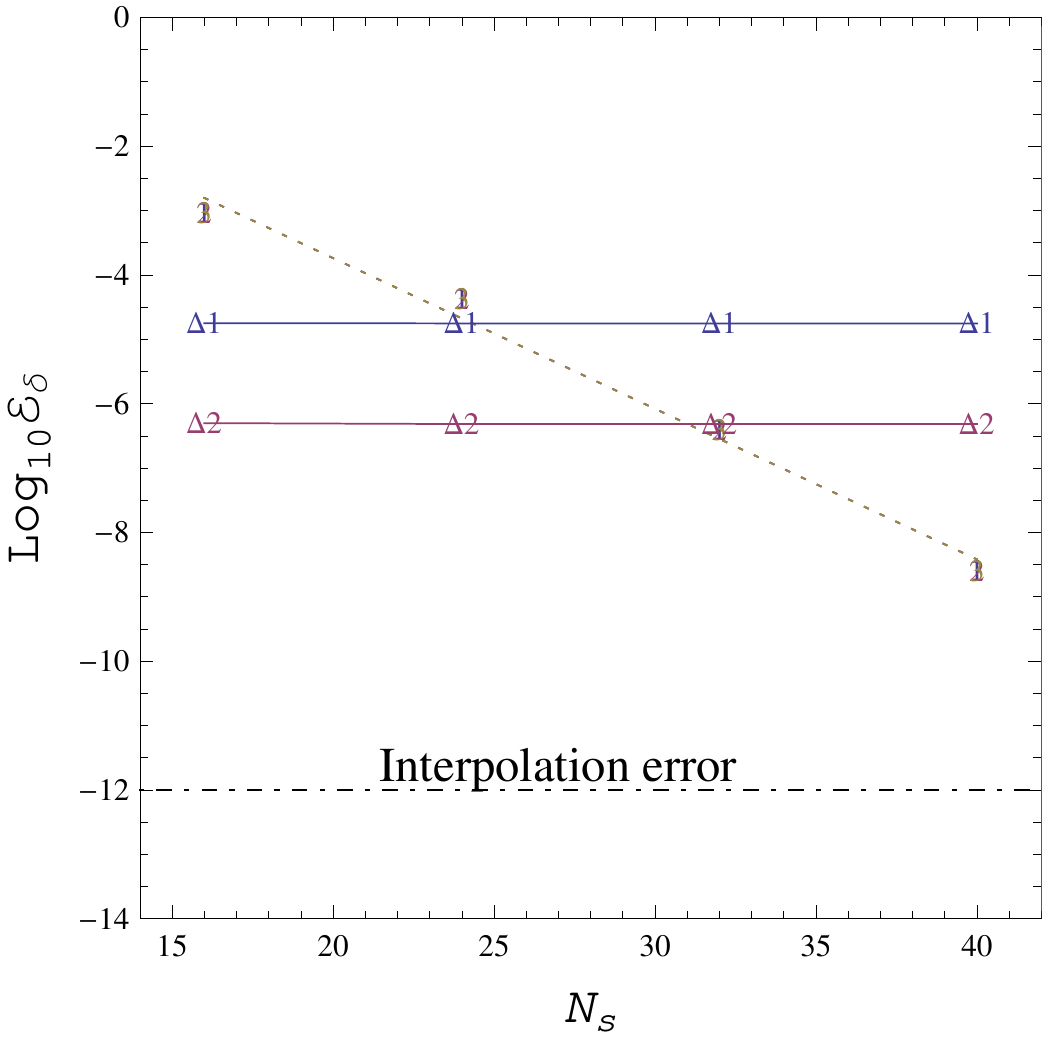} 
\caption{Convergence of the solution as the size of the spatial
grid or the Lagrangian order is increased. Cauchy differences of
density on the OBS Lagrangian grid are plotted.
The dotted line traces the differences for calculations
having increasing numbers of
grid points ($N_s$ and $N_s+8$) at fixed Lagrangian
orders ($n=1$, $2$ and $3$). In fact, the calculated
differences, labeled by $n$, all overlap.
The straight dotted line indicates the exponential convergence
expected for a spectral method.
The solid, horizontal lines
are Cauchy differences with respect to Lagrangian
order ($\Delta$n refers to orders $n$ and $n+1$) at fixed $N_s$.
Both types of errors are generally present.
For small grids the grid-related spatial errors
dominate and for large grids the Lagrangian-related finite order
errors dominate.
The crossover from $N_s$-dominated to $n$-dominated errors
occurs at $N_s^*$. Additional calculations (not shown)
demonstrate that $N_s^*$ is smaller for smoother initial data.
All the errors shown are
much larger than the interpolation error shown by the dashed-dotted
line. }

\label{smoothtophat}
\end{figure}

Runs with $16 \le N_s \le 48$ and $1 \le n \le 3$ were carried
out for fixed smoothing.  We evaluated the Cauchy differences
as the grid was refined (${\cal E}(N_s;N_s+8)$ at fixed
$n$) and then as the Lagrangian order was increased (${\cal E}(n;n+1)$
at fixed $N_s$) for $q=Y$ (on the OBS Lagrangian grid).
\capfigref{smoothtophat} shows the results for the density. 
The Cauchy differences for varying
grids are virtually identical for $n=1 - 3$ (this is not
obvious from the figure, because the lines for different $n$
overlap). The Cauchy differences for varying Lagrangian do not depend
upon $N_s$ (this is obvious from the fact the Cauchy
differences are basically horizontal lines). Define the grid
crossover $N_s^*(n)$ to be the value of $N_s$ that
satisfies ${\cal E}(\{n,N_s\}; \{n,N_s+8\}) =
{\cal E}(\{n,N_s\};\{n+1,N_s\})$ for a given $n$. 
The behaviour of the formal measure of convergence is
\beq
{\cal E}(\{n,N_s\};\{n+1,N_s+8\}) \sim \left\{
\begin{array}{cc}
{\cal E}(\{n,N_s\}; \{n,N_s+8\}) & {\rm if\ } N_s < N_s^* \\
{\cal E}(\{n,N_s\}; \{n+1,N_s\}) & {\rm if\ } N_s > N_s^*
\end{array}
\right.  
\eeq 
The grid-related refinement dominates in the first case and
order-related refinement in the second. The figure makes
clear that the spatial differences decline exponentially
with $N_s$ (as they should for a spectral method) and the
analytic structure for the LPT expansion suggests that order
errors decline exponentially with $n$ (this will be tested
more directly in a succeeding section). One surmises that
for two sources of 
exponential errors, the transition $N_s^*$ should be a linear function of
$n$.

The specifics of the transition from grid to order-dominated
convergence are problem dependent. We have investigated
initial conditions which differ from those above only in
the degree of smoothing used (choice of $\sigma$). We
find $N_s^*$ is larger when initial data is less smooth and
vice-versa.
 
The practical upshot is that if the solution is smooth and
$N_s>N_s^*$ then the spatial errors are much smaller than
changes associated with incrementing the Lagrangian order or
decreasing the step size over finite ranges of $n$ and
$N_t$, respectively. We attempt to conduct all the rest of
our tests in the limit $N_s>N_s^*$ when the grid errors are
negligible.

\subsection{Convergence with Lagrangian order $n$ and number of steps $N_t$ }
\label{sec:ordersize}

This section presents tests of convergence with Lagrangian
order $n$ and number of steps $N_t$ for initial conditions
arising from a random Gaussian field at fixed spatial grid
size. Two types of initial data are considered: Zel'dovich
initial conditions (velocity proportional to the
acceleration field; irrotational flow) and generic initial
conditions (initial velocity chosen arbitrarily from random
Gaussian fields; transverse and longitudinal components
present and of comparable size). We focus on the truncation
error for Lagrangian order $n$ and re-expansion frequency
$N_t$. We choose grids, initial conditions and time
intervals for evolution for which we can be reasonably sure that
grid-related errors at fixed $N_s$ are so small that they do
not interfere with the Cauchy differences for varying $n$
and $N_t$.

We proceed as follows. The power in the initial data is
limited to frequencies less than half the Nyquist
frequency. In this discussion, we refer to wavenumbers as `frequency' ($k = 2 \pi/X$), not to be confused with the re-expansion frequency $N_t$. 
For box-length $L$ and grid size $N_s$, the magnitude of the Nyquist frequency is $\pi N_s/L$.  
Since gravitational dynamics is intrinsically
non-linear we know that the power in these initially zeroed
modes will grow as collapse proceeds.  Ultimately, any power
that reaches or exceeds the Nyquist frequency will manifest
as error since the Nyquist frequency is fixed because the
grid is fixed. The size of
the power in the vicinity of the Nyquist frequency is taken
to be a surrogate for error incurred by using a finite grid.
We monitor the power that builds up to make sure that the
spatial errors remain negligible.

\def\spsum{{\sum_{\left({\bf k}\right)}}}
The power in the Nyquist frequencies is
\beq
P_{f,Nyq} =
\sqrt{  \frac{1}{N_s^3} \spsum |{\tilde f}^{N_t}({\bf k}) |^2 } 
\eeq
where ${\left({\bf k}\right)}$ is the set of wave numbers having a
Cartesian component equal to the Nyquist frequency (or, we
can monitor the power within a given range of the Nyquist
value).
Consider, for example, the Cauchy difference with respect to
$N_t$ on the OBS Lagrangian grid. This may be rewritten using Parseval's theorem (\citealt{numrecipes})
\beq
{\cal E}_f (N_t;N_t\phantom{}^{'}) = \left (\frac{1}{N_s^3} \sum_{\bf X} | f^{N_t\phantom{}^{'}}({\bf X}) - f^{N_t}({\bf X})|^2 \right)^{1/2}=   \left (\frac{1}{N_s^3}\sum_{\bf k}| {\tilde f}^{N_t\phantom{}^{'}}({\bf k})- {\tilde f}^{N_t}({\bf k})|^2 \right)^{1/2}, 
\eeq 
where ${\tilde f}$ is the Fourier transform of $f$ defined as ${\tilde f}_m = N_s^{-1/2} \sum_{l=1}^{N_s} f_l e^{2\pi i (m-1)(l-1)/N_s} $.
The maximum contribution to the total that comes from the Nyquist modes is 
\beq 
\sqrt{ \frac{1}{N_s^3} \spsum |{\tilde f}^{N_t\phantom{}^{'}}({\bf k})  - {\tilde f}^{N_t}({\bf k}) |^2 }\lta 2 \sqrt{  \frac{1}{N_s^3} 
\spsum |{\tilde f}^{N_t}({\bf k}) |^2 } = 2 P_{f,Nyq}.
\eeq
where the Nyquist components for
$N_t$ and $N_t\phantom{}^{'}$ are estimated to 
be of the same order of magnitude. When
${\cal E}_f(N_t;N_t\phantom{}^{'}) >> P_{f,Nyq}$ we infer that the
uncalculated finite grid errors are small compared to the
calculated Cauchy differences.  Likewise, when
${\cal E}_f(n;n') >> P_{f,Nyq}$ we infer grid errors are small
compared to the Cauchy differences for $n$.
The fractional overdensity and peculiar velocity are not independent.
We always use $\max_{f=\{\delta,v\}} P_{f,Nyq} $ in the figures that follow.

\subsubsection{Zel'dovich initial conditions}
\label{sec:cosmocase}

The initial density is chosen to be a specific realization
of a random Gaussian field and the initial peculiar velocity
is set to be proportional to the acceleration at
each point (Appendix \ref{App:zeldovich}). The initial
conditions satisfy ${\bf v}_c(t_0)=0$.
We calculated evolution over the same time
interval for Lagrangian orders $n = 1-3$ and
frequency of re-expansion $N_t=1-5$.

Before presenting the convergence results let us first
discuss an important feature of this flow. The velocity is
irrotational at the initial time in the observer
frame. Kelvin's theorem states that for an ideal fluid under the
influence of a conservative force, the velocity circulation
around a closed contour comoving with the fluid stays
constant in time. As a consequence,
the vorticity of the velocity field is conserved. In
particular, if the fluid is irrotational initially it stays
irrotational (\citealt{landaulifschitz}; EB97).

Kelvin's theorem applies to the exact dynamics in the
Eulerian coordinate system. Although the initial conditions are irrotational
in both Eulerian and Lagrangian systems, the Lagrangian
calculations do not preserve irrotational flows in the
Lagrangian coordinate system because there is no Lagrangian
version of Kelvin's theorem. If the Lagrangian calculation were exact
then the flow would satisfy the exact conservation law for circulation after
transformation back to the Eulerian coordinate system.

Our calculation is a finite order Lagrangian calculation
with finite step size. We expect the exact Eulerian result
that the flow should be irrotational (at any time) will
emerge only as convergence is achieved. All non-zero
transverse components in the OBS frame are errors and should
necessarily diminish as the calculation is refined.

\begin{figure}
\includegraphics[width=6cm]{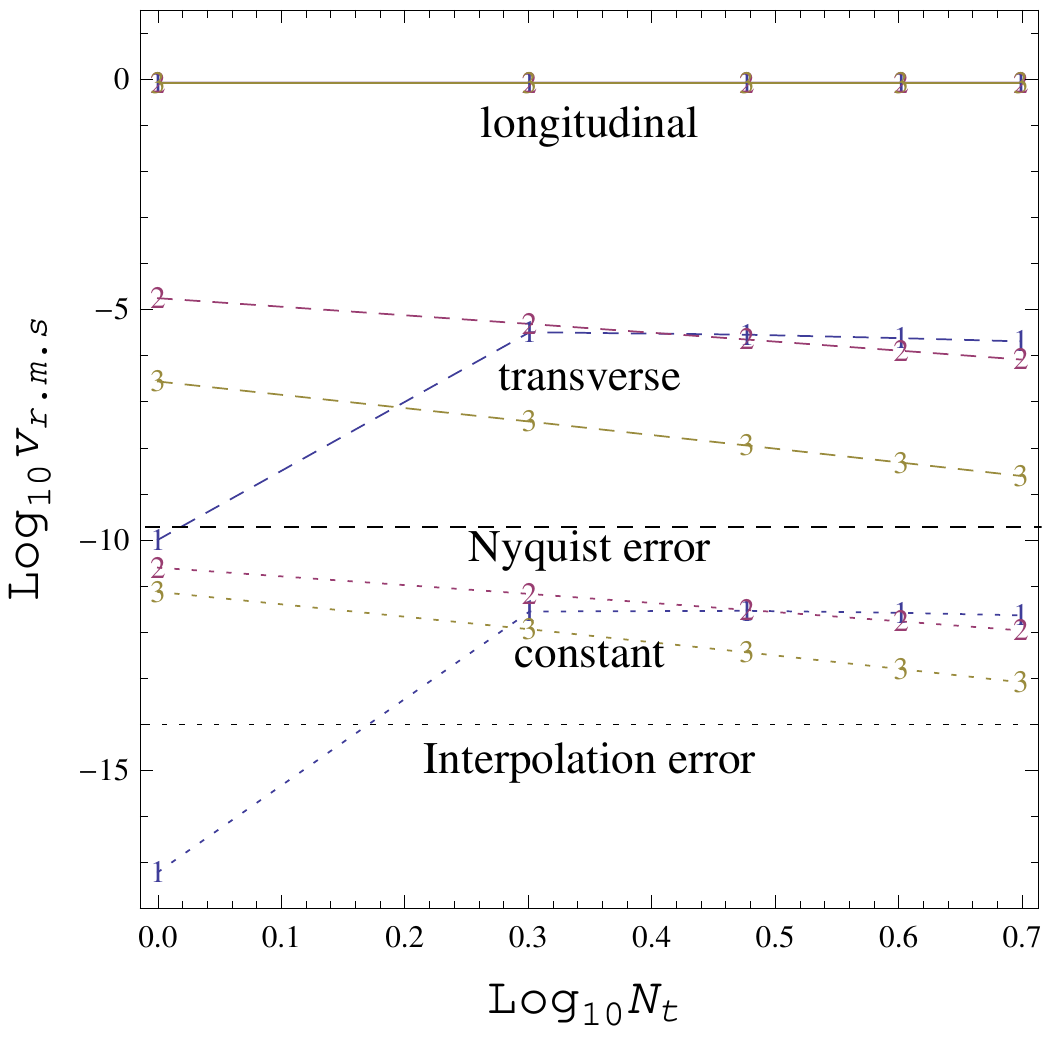} 
\caption{Root mean square of ${\bf v}^{(L)}$ and ${\bf
    v}^{(T)}$ and ${\bf v}_c$ at the final time in the
  observer frame for the Zel'dovich initial conditions as a
  function of $N_t$ the number of steps taken. Results for
  Lagrangian orders $n=1-3$ have different colours and plot
  markers. Note that transverse components converge to zero
  in agreement with Kelvin's theorem.}
\label{strengths}
\end{figure}

\capfigref{strengths}
summarizes the scale of different velocity components at the
final time in the OBS frame as a function of $N_t$ for
calculations at Lagrangian orders $n=1-3$. In all cases the
longitudinal component dominates and the transverse term is
small by comparison. The transverse term decreases as $N_t$
increases and as $n$ increases. Both these trends show that
solution behaves qualitatively in accord
with Kelvin's theorem.

Three other important scales are indicated on
\figref{strengths}.  The interpolation error is very small,
consistent with the effect of machine precision; it is
inconsequential for all our considerations.  The Nyquist
error provides an estimate of the error associated with the
spatial grid. It is considerably larger than interpolation
error but less than the scale of the transverse
velocities mentioned above. Finally, the mean
peculiar velocity ${\bf v}_c$ (labeled ``constant'' because
it does not depend upon position) is
shown. As discussed in \S \ref{sec:Pcons} ${\bf v_c}$ is
generally non-zero.  In this run, the Zel'dovich initial
conditions dictate ${\bf v}_{c,0}=0$ and the time interval
of evolution is small. Hence, the final ${\bf v_c}$ is small in
the sense of being no bigger than the Nyquist
error. We infer that frame shifts play no essential role in
these particular results.

\begin{figure}
\includegraphics[width=16cm]{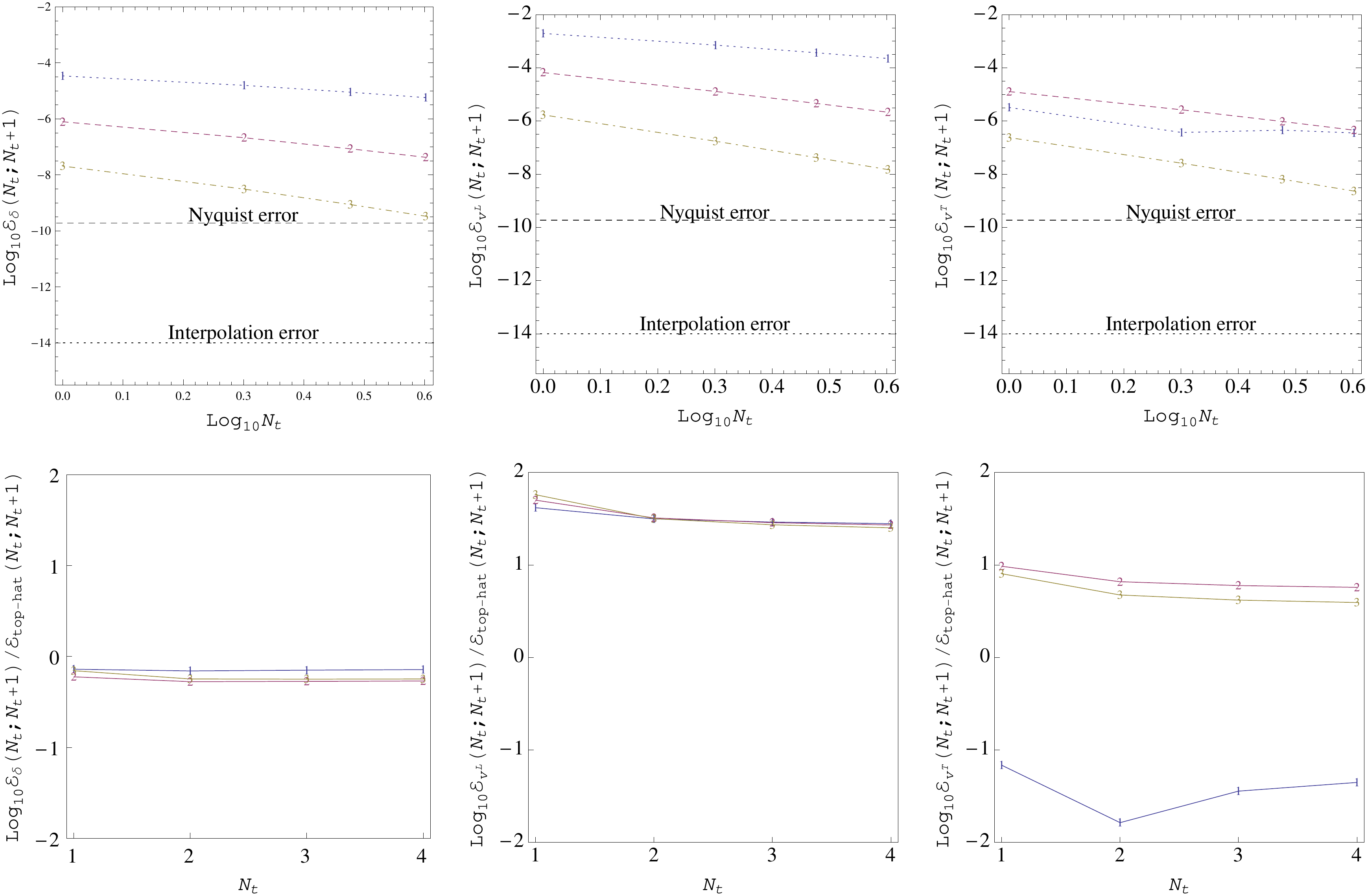} 
\caption{Convergence with respect to frequency of
  re-expansion for Zel'dovich initial conditions. Top panel:
  Cauchy differences at the final time 
${\cal E}_\delta$, 
${\cal E}_{{\rm v^L}}$, 
${\cal E}_{\rm v^T}$ 
  (comparing
  ${N_t,N_t+1}$ 
  for fractional overdensity,
  longitudinal peculiar velocity and transverse peculiar
  velocity, respectively). Nyquist error and interpolation
  error levels lie below those of the Cauchy differences and do
  not interfere. For all
  variables, at any order, the error decreases as the $N_t$
  increases. The rate of convergence is better for larger
  Lagrangian order $n$. Bottom panel: the ratio of Cauchy differences in
  the inhomogeneous calculation to those
  of the top-hat. It is noteworthy that the ratio is roughly constant for
  larger $N_t$ and $n$. This provides a confirmation that
  the theoretical top-hat convergence rate describes the inhomogeneous
  case.
}
\label{multistep_cosmo}
\end{figure}

Now consider the Cauchy differences for the solutions in
\figref{multistep_cosmo}. The upper panel presents the
results for density, longitudinal and transverse components
of velocity (left to right) in the OBS Eulerian frame
as the number of steps $N_t$ changes. Lines are
labeled by Lagrangian order $n$. The Nyquist and
interpolation errors lie below the Cauchy differences
indicating that the truncation errors for order and frequency
of re-expansion have been isolated from the effects of grid
size and numerical precision. The slope of the lines slightly increases with higher $n$ indicating faster convergence.

For the top-hat it was found theoretically, and confirmed
empirically, that for a small time step $\Delta t/t_0<<1$
the Cauchy differences decreased with $N_t$ and $n$ as
\bea
{\cal E}_{top-hat}(N_t;N_t+1) &  \sim & |g_{N_t+1,n} - g_{N_t,n} | \\
{\cal E}_{top-hat}(n;n+1) & \sim & |g_{N_t,n} |,
\label{tophatscaling}
\eea
where 
\bea 
g_{N_t,n} &=& K_{N_t,n} \cos\theta \sin^n\theta \Delta^{n+1} \left(\frac{\Delta t}{t_0}\right)^{n+2}\\
K_{N_t,n} &= &\frac{1}{9} \left( \frac{-2}{3}
\right)^n \left( \frac{ N_t - \frac{n}{2+n} }{N_t^{n+1}}
\right).  
\eea 
The quantities $\Delta$ and $\theta$ characterize the
strengths of the initial density and velocity
perturbations ($\delta =  \Delta \cos \theta$ and $\delta_v =  \Delta \sin \theta $). Note that $g_{N_t,n}$ scales
exponentially with Lagrangian order $n$ for any given $N_t$. This is expected because the
$n$-th term in the LPT series is $\propto \epsilon^n$, where
$\epsilon$ is the magnitude of the initial
perturbation. Conversely, the theoretical error scales as a
power of the number of steps $\propto N_t^{-n}$ as $N_t$
increases.  The latter also implies that the slope steepens
with Lagrangian order $n$.

The lower panel of \figref{multistep_cosmo} presents a
quantitative comparison of the top-hat convergence rates 
to the more general problem that has been calculated. The
figure shows the ratio of ${\cal E}$ to that deduced
previously for the top-hat. (For Zel'dovich initial
conditions $\theta = \arctan(-1/3)$ and $\Delta
=\sqrt{10} \delta /3$; we used the root mean square value for $\delta$).
The ratio is approximately constant implying that
the convergence rates with respect to $n$ and $N_t$ for
inhomogeneous initial conditions are close to those of the
top-hat. The agreement is very good for the density and
longitudinal velocity.  The only anomaly appears to be
related to the relative size of the first and second order
transverse velocity differences (far right panel). The
second and third order terms scale as predicted theoretically.

The explanation is enlightening. One might first imagine
that the cause was related to the fact that the top-hat
collapse does not include any components of transverse
velocity. Then it would be unsurprising if the theoretical
result failed to capture the apparent inversion in
transverse components. But that is not what we
find. The key point is already present in
\figref{strengths} which shows that the magnitude of the
transverse velocity is very small after a single, first
order step but much larger after a single, second order step
or after multiple steps of first order.

We find the explanation is related to the way in which the
Lagrangian system ``recovers'' the result implied by
Kelvin's theorem, i.e. that the transverse component in the
Eulerian system should ultimately vanish. When the initial conditions satisfy velocity proportional to acceleration, the lowest order, non-vanishing contribution to the Lagrangian frame transverse velocity is third order (BE93; B94). However, the transformation from Lagrangian to
Eulerian coordinates is essential, i.e. the transformation
itself can contribute to the Eulerian transverse 
velocity. The Eulerian transverse velocity (defined as ${\bf w}_E({\bf r}, t) = \nabla_{r} \times {\bf v}_{OBS}$) is equal to the Lagrangian transverse velocity (defined as ${\bf w}_L({\bf Y}, t) = \nabla_{Y} \times {\bf v}_{OBS}$) plus `extra' terms that are quadratic and cubic combinations of 
the displacement vector and its time derivative (see Appendix \ref{app:vTgenerate}) 
\beq 
 {\bf w}_E({\bf Y}, t)  = {\bf w}_L({\bf Y}, t) + {\rm Terms}[{\bf p}, {\dot {\bf p}}] + {\rm Terms}[{\bf p}, {\bf p}, {\dot {\bf p}}].
 \label{vorticity}
\eeq
When the calculation is done with a first order, single step LPT scheme for Zel'dovich initial conditions, these `extra' terms vanish exactly and we get both ${\bf w}_L({\bf Y}, t)=0$ {\it and}  ${\bf w}_E({\bf Y}, t) =0$ at all times. 
When the calculation is done with a second order, single step LPT scheme for Zel'dovich initial conditions, ${\bf w}_L({\bf Y}, t)=0$ at all times, but the `extra' non-zero terms are third and higher order in the expansion parameter (`extra' terms that are second order vanish). 
\beq 
 {\bf w}_E({\bf Y}, t)  = {\bf w}_L({\bf Y}, t) + \mathcal{O}(\epsilon^3)
\eeq
${\bf w}_E({\bf Y}, t) \neq 0$ except for the initial time. So, we have the unexpected consequence that a
single step of a first order LPT calculation will yield zero
Eulerian transverse velocity in the OBS frame but that a single step
of a higher order calculation will generally yield a
 non-zero Eulerian transverse velocity. Both the first order and the
infinite order calculations will satisfy Kelvin's
theorem. Of course, the first order solution is not exact
and we can expect convergence only as the order increases. In general, with a $n$-th order calculation, one expects ${\bf w}_E({\bf r}, t) = \mathcal{O}(\epsilon^{n+1})$ which relies on complicated cancellations between all terms of order less than $n$ between the three pieces on the r.h.s. of \eqnref{vorticity}.  

\capfigref{strengths} shows that multiple steps of a first
order scheme builds up a non-zero transverse velocity.  This
might appear to be at odds with the discussion above in
which a single, first order step exactly satisfies Kelvin's
theorem. The reason is that even one step will generate a
state where the initial proportionality between velocity and
acceleration is broken.  In subsequent steps even a first order LPT scheme can then give 
non-vanishing contribution to transverse velocity in
the Eulerian frame (due to the second and higher order terms in the transformation between Eulerian and Lagrangian coordinates). 
Multi-step, first order
calculations may generate transverse velocity. Convergence and
consistency with Kelvin's theorem is expected as the number
of steps increase.

Ultimately, only the asymptotics matter: we have shown that
refinements in order and/or step size both lead to smaller
transverse motions in the Eulerian frame.  Higher order
Lagrangian schemes converge more quickly in the asymptotic
limit. For example, in the top right hand panel the line for
$n=3$ lies below that for $n=2$ and its downward slope (as a
function of $N_t$) is greater.

We conclude that the LPT re-expansion converges for
Zel'dovich initial conditions and, moreover, that it does so
at a rate which is essentially identical to that
theoretically derived in the case of the top-hat.

\subsubsection{Generic initial conditions}
\label{sec:gencase}
\begin{figure}
\includegraphics[width=16cm]{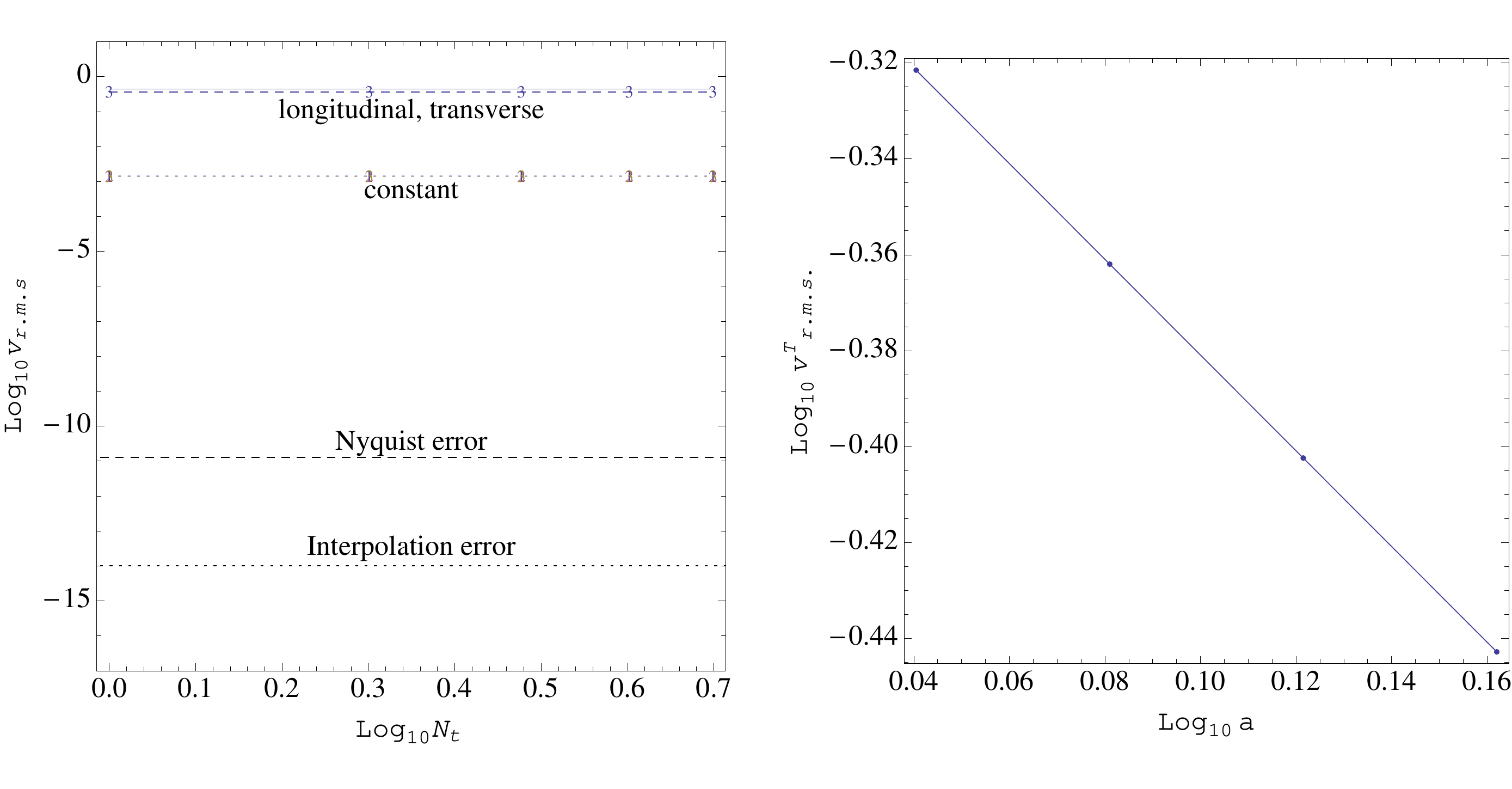}
\caption{Root mean square of various components of the final
  velocity in the observer frame (left) and the transverse
  velocity as a function of the scale factor (right).  The
  longitudinal and transverse components are roughly the
  same magnitude initially and, on account of the
  small time advanced, finally. However, on an expanded scale
  magnitude of the transverse velocity decreases $\propto 1/a$. Dots are
  data, line is the fit. The theoretically expected slope is
  $-1$ and best fit line has slope $-0.99$. All orders have
  similar behaviour; first order data is shown here. }
\label{genstr}
\end{figure}

In this test the initial density and peculiar velocity are
independent realizations of a Gaussian random field. The
latter has longitudinal and transverse components of
comparable magnitude. We follow the evolution of the initial
conditions for varying Lagrangian order ($n=1-3$) and
re-expansion frequency ($N_t=1-6$). \capfigref{genstr}
summarizes (left panel) the root mean square velocity at the
final time and (right panel) the evolution of the transverse
velocity.  For the small time step considered here the
longitudinal and transverse parts are still comparable at
the final time but (right panel) the transverse mode decays
$\propto 1/a$. Here, ${\bf v}_c$ is greater than in the
Zel'dovich case and points to the importance of frame shifts. 

\begin{figure}
\includegraphics[width=16cm]{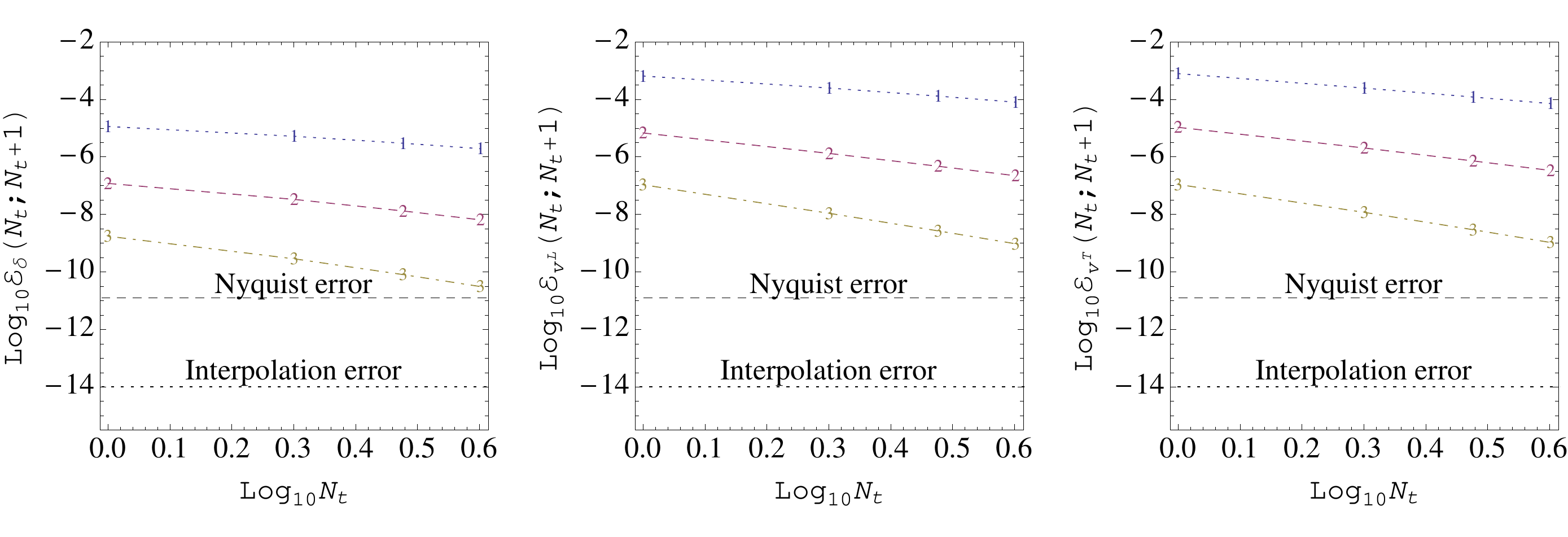} 
\caption{General initial conditions: convergence with
  respect to Lagrangian order and frequency of
  re-expansions. The coding of the graph is the same as
  \figref{multistep_cosmo}.  Cauchy differences for $N_t$
  decrease as $N_t$ increases; the rate of convergence
  improves for higher Lagrangian order $n$. The scaling of
  these errors with respect to a top-hat is similar to the
  lower panel of \figref{multistep_cosmo} and is not shown
  separately here.  }
\label{fig:gencase}
\end{figure}

\capfigref{fig:gencase} shows the Cauchy differences for
changes in $N_t$. The differences decrease with increase in
$N_t$ and Lagrangian order $n$. The transverse term here
behaves in a qualitatively identical manner to the
longitudinal case (there is no low-order anomaly like that
seen in the Zel'dovich case). The scaling of these errors
closely follows the trend established for the top-hat (the graphs
are not repeated here).

Our main conclusion drawn from the results for the generic
test case is that the full gravitational dynamics has been
successfully implemented in the LPT re-expansion method.

\subsubsection{Frame shifts again} 
\label{Pconstest}

We had previously illustrated frame shifts in the simplest
possible case, one-dimensional problems. Here we revisit the
issue in the context of the general case just discussed. We will show that the numerical
convergence as measured by the Cauchy differences is
destroyed and momentum conservation broken when frame
shifts are omitted. 

\begin{figure}
\includegraphics[width=16cm]{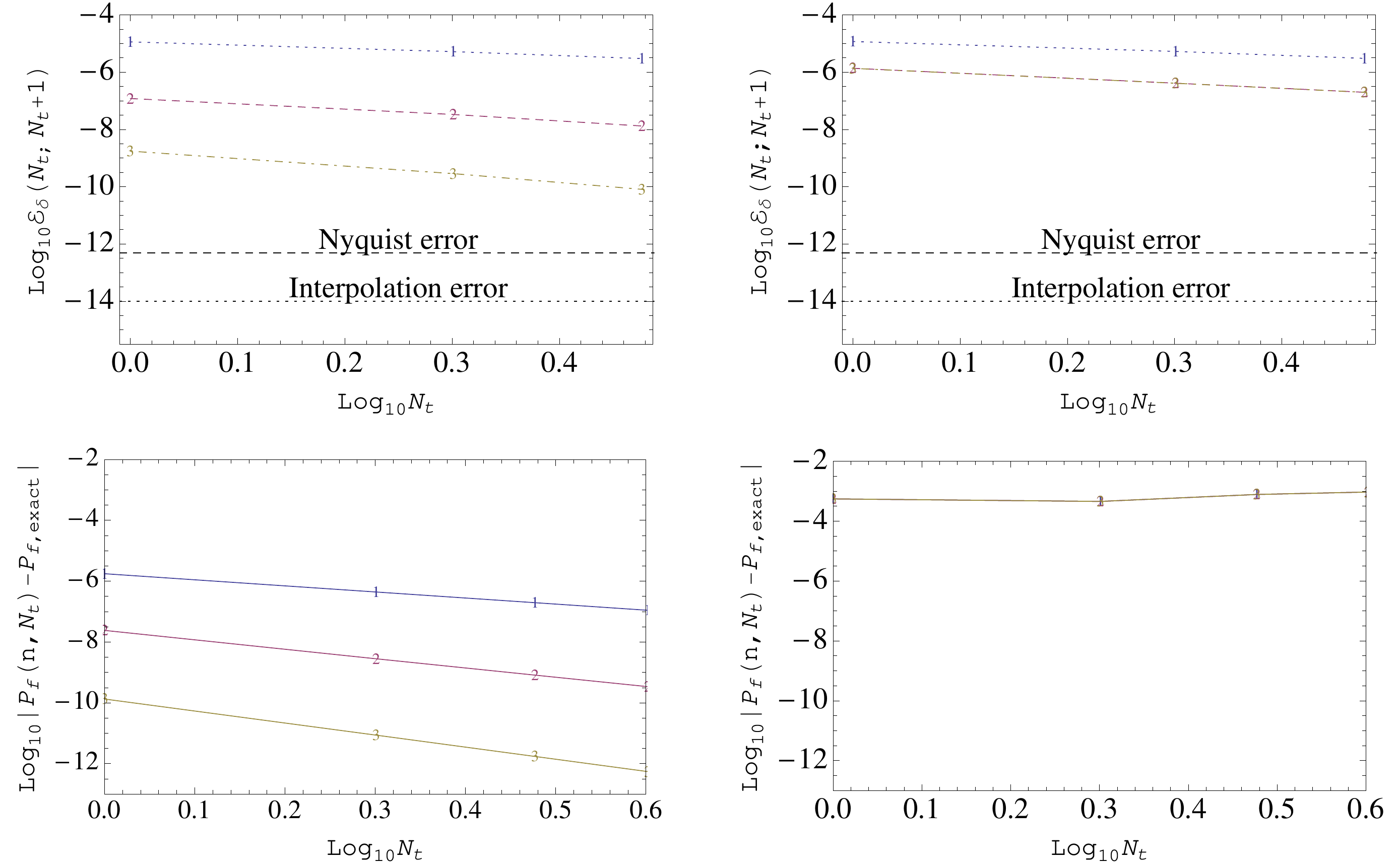} 
\caption{Frame shifts and momentum conservation (general
  case). Top panel: Cauchy differences for density with (left) and
  without (right) frame shifts. Bottom panel: momentum
  conservation errors with (left) and without (right) frame shifts. Frame shifts are crucial to ensure convergence with Lagrangian order and ignoring them leads to a breakdown of momentum conservation. }
\label{comparegen}
\end{figure}

\capfigrefs{comparegen} compares convergence of the density
field in calculations with (left) and without (right) frame
shifts. The left panel shows the same sort of scaling with
$n$ and $N_t$ as seen in the two previous
sections. Increasing the Lagrangian order and/or increasing
the frequency of re-expansion reduces the differences. The
right panel shows an anomaly, that the Cauchy differences
for $n=2$ and $n=3$ lie on top of each other. This implies
that convergence with Lagrangian order is broken.

The lower panel compares momentum conservation error with
(left) and without (right) frame shifts. 
The momentum, as defined in \eqnref{Pdefn}, is
calculated at initial and final times, ${\bf P}(t_0)$ and
${\bf P}_f$, respectively. The former is known exactly from
the choice of initial conditions. After the evolution ${\bf
  P}_f$ is known numerically and it may be compared to the
exact result ${\bf P}_{f,exact} = a(t_0) {\bf
  P}(t_0)/a_f$. We calculated the error as the absolute
difference between the exact and approximate value summed in quadrature over the three directions.
The figure (left) shows that conservation is
achieved asymptotically with frame shifts. By contrast, there
is no evidence for improvement when the frame shifts are
omitted (right). Increasing the Lagrangian order and/or
increasing the frequency of re-expansion reduces the error.

\subsection{Summary of convergence tests}

The tests indicate that the LPT re-expansion method works in
principle and in practice for inhomogeneous initial
conditions. This is the natural extension and generalization
of our previous work on the top-hat model. These tests probe
convergence when various combinations of grid size,
Lagrangian order and/or re-expansion frequency are
increased. Testing the rates separately has yielded results
consistent with the underlying solution methodology
(exponential in number of grid points for the spectral
representation of spatial functions, exponential in
Lagrangian order for the perturbation expansion, geometric
in step size at fixed Lagrangian order).  An essential part
of the method is frame shifts.  Ignoring them leads to
violations of momentum conservation and breaks the
theoretically expected convergence rates.

\section{Conclusion}
 \label{sec:conclusion}

We have developed a new method to calculate the growth of
large scale structure based on Lagrangian perturbation
theory. It is designed to handle generic
inhomogeneous perturbations on sub-horizon scales having
non-relativistic velocities and weak gravitational fields.
It can follow the system evolution until the formation
of the first caustic.

Many of the fundamental ideas were first discussed in the
context of spherical top-hat cosmologies in NC11. There we
demonstrated the existence of limitations on the convergence
of the series expansion which is the basis of LPT. We showed
that even if one could work to infinite order in the LPT
expansion parameter there are bounds in time that limit how
far forward a single step can extend.  Such limitations are
more restrictive than the onset of particle crossings. To
circumvent the bounds we introduced the idea of re-expanding
the solution in overlapping time domains (``LPT
re-expansion''). Our work generalized LPT from a single-step to
a multi-step scheme permitting evolution all the way up to
the onset of multi-streamed flow.

This paper extends the analysis from spherical top-hat
cosmologies to generic initial conditions in the context of
Newtonian cosmology. A new and important theoretical point
is the existence of frame shifts. Frame shifts originate
because the starting point of traditional derivations of the
LPT series involves taking derivatives of the geodesic
equation or its equivalent. The perturbative theory and
solution obtained in this way is insensitive to 
time-dependent translations and will generally not satisfy
the original equations of motion unless this freedom is
fixed (no such freedom was present in the spherical cases
studied in NC11). We introduce two conceptual frames to
explore the situation. The OBS frame is the frame in which
the physical initial conditions are specified and the answer
sought. The EB frame is the frame in which the perturbative
solution is obtained via the formal LPT expansion described
in EB97. We describe in detail how to transform from one
frame to the other. The relationship between OBS and EB
frames is non-trivial. For example, it is not set by a physically
conserved quantity. In many instances frame shifts are
required, e.g. in a finite cosmology simulation, the EB
frame will generically differ from that of the preferred OBS
frame tied to the cosmic microwave background.  At each step
one must determine the relationship anew.  We argued that
ignoring the frame shifts generically leads to a breakdown
in momentum conservation.

The multi-step algorithm we have developed can be summarized
as follows. Start with initial density and velocity defined
in the OBS frame with Lagrangian coordinates equal to
uniform Eulerian grid coordinates at time $t$. Transform
quantities to the EB frame (the density remains the same,
velocity differs by a constant).  Use the EB formalism to
advance the density and velocity in the EB frame from $t$ to
time $t+\delta t$ subject to the restriction imposed by the
time of validity, the span over which the LPT expansion
converges.  Compute the frame shifts accumulated during the
evolution and transform the EB solution back to the OBS
frame at the end of the step.  Interpolate the final OBS
quantities onto the uniform Eulerian grid at time $t+\delta
t$. These quantities serve as the initial data with new
initial Lagrangian coordinates equal to Eulerian coordinate
for the next step.

The algorithm was numerically implemented and tested. All
tests were done in a $\Omega_m=1$ cosmology over time
intervals that obeyed the predicted time of validity. Since
these are meant to be `proof of principle' tests we
performed them on grids of small size. 

Some tests involve comparisons to exact answers.  The
solution for a simple, plane-parallel problem can be written
down without reference to LPT. Since first order LPT should
produce an exact result this enables several different
checks to be carried out. We showed that our first order LPT
treatment with frame shifts matches the exact solution. We
also established that ignoring the frame shifts gives rise
to problems and inconsistencies.  Several different lines of
argument (involving momentum conservation, convergence,
single versus multiple steps) substantiated the fundamental
role of the frame shift even in the context of this
simplest-of-all problems. Discrepancies introduced by
ignoring frame shifts might go undetected for some specific
initial conditions or in a single step scheme but they
become clearly apparent even in this simple geometry for general 
initial data and/or while utilizing a multi-step scheme.

As a second exact comparison we tested our results against
the analytic example given in 
\citet{buchert_performance_1997}. This allows a check of the
implementation of the higher order LPT terms. This 
provided a complete check of the non-zero (spatial and
temporal) terms up to third order and a consistency check
that the spatial parts of all the other terms vanish. The
specific numbers of each type of term are given in the text.
The frame shift, which vanishes for this particular example,
is also checked by this test.

Other tests involve situations where the exact answer is
unknown and we focus on whether the convergence of the
algorithm agreed with theoretical expectations.  There are
three control parameters: grid size $N_s$, Lagrangian order
$n$ and the frequency of re-expansion $N_t$. We first discuss
theoretical convergence to the exact answer. Generally,
$N_s$ and one (or both) of $n$ and $N_t$ must increase
simultaneously.  The refinement must be made subject to
the time of validity criterion.

For diagnostic purposes we show it is possible to choose
initial conditions and parameters such that errors are
largely set by a single parameter. This allows us to
study the nature of the convergence of that part of the
calculation. To demonstrate convergence with grid size we
represented the initial density and velocity with a
spherical top-hat profile filtered with a Gaussian. We
showed that for these smooth initial conditions, the errors
due to the finite grid size decreased exponentially with
$N_s$, when other errors due to Lagrangian order were
negligible.  This was in accordance with the expected
behaviour because spectral methods are used to solve the
spatial equations. Likewise, when the grid size errors are
very small the convergence with Lagrangian order is
exponential in accordance with the basic understanding that
LPT is a power series expansion in the perturbation
parameter. We show that there is a sharp transition from
convergence being controlled by grid size $N_s$ to it being
controlled by Lagrangian order $n$.

For the remaining studies, we set $N_s$ large enough that we
could focus on the convergence with Lagrangian order $n$ and/or
the number of time steps $N_t$. We considered initial data
based on a random Gaussian field.  Two types of initial
conditions were considered: Zel'dovich initial conditions,
where the velocity was proportional to the acceleration
field, and generic initial conditions, where the initial
velocity includes both longitudinal and transverse
components. Though unanticipated we discovered that the rate
of convergence of these complicated examples scaled nearly
identically as the rate of convergence for the simple
top-hat cosmologies (whose rates were analytically derived
and numerically confirmed in NC11).  To summarize, the
convergence with $n$ is exponential and that with $N_t$ is
algebraic. The internal consistency for solutions achieved
when varying $n$ and $N_t$ (a common asymptotic solution and
Cauchy differences which diminish at the theoretically
anticipated rate) provides strong evidence that the system
is being solved correctly.  We also demonstrated that when
the frame shifts are omitted the convergence tests fail
and conservation of momentum is violated.

We explored how Kelvin's theorem is satisfied for
situations in which the initial Eulerian vorticity vanishes.
Kelvin's theorem implies that the Eulerian vorticity
vanishes at all future times prior to the formation of
caustics. Since this is an exact result it serves as a
useful check on the end-to-end solution, i.e. the
transformations between OBS and EB frames and the LPT
calculation within the EB frame. We confirmed numerically
that the magnitude of the Eulerian vorticity converges to
zero in the asymptotic limit just as it should. This is a
non-trivial check as can be seen from the following
considerations. If the initial vorticity vanishes then
vorticity as measured in the Lagrangian frame is generated
at third and higher orders (for single step Lagrangian
solution; BE93 and B94). What happens to
these contributions in the OBS frame? They should vanish as
the calculation order increases. The key is that vorticity
as measured in the Lagrangian frame is not the same as
vorticity as measured in the Eulerian frame. This can be
seen in several ways. For example, a second order LPT
calculation for the initial conditions described above
yields zero vorticity in the Lagrangian frame (to second
order) but this still gives rise to a non-zero vorticity in
the Eulerian frame because of the transformation between the
two frames. Alternatively, in a multi-step scheme, Eulerian
vorticity is generated even with a first order LPT
scheme. The appearance of Eulerian vorticity at a given finite
order does not violate the Kelvin circulation theorem as
long as the order of that contribution increases with the order of the
calculation itself. This is the situation that is indicated
by the numerical results. (Likewise, increasing the number of
steps causes the magnitude of the Eulerian vorticity
to diminish.) Prior studies have examined the density and/or the Lagrangian displacement
and, to the best of our knowledge, these aspects of the
solution have never been tested. The frame transformation
is needed to achieve this consistency.

We now consider the future utility of this new method.

The choice of the exact number of time steps, Lagrangian
order and grid size which govern the numerical errors will
ultimately depend on the application at hand.  The numerical
errors can be made as small as desired.  However, we have
not addressed the important drawback of LPT, which is its
inability to model physics beyond shell crossing. Real
cosmological applications will be limited by the occurrence
of multi-streaming. Any approximation to account for the
velocity dispersion (\citealt{adler_lagrangian_1999}; 
\citealt*{buchert_extending_1999}; \citealt{morita_extending_2001}; \citealt{tatekawa_third-order_2005-1}) or alterations to the basic
dynamics (such as the adhesion approximation; see reviews
\citealt{sahni_approximation_1995}; \citealt*{gurbatov_large-scale_2012}) will
introduce physical errors as opposed to the numerical errors
with which we have dealt. The
algorithm presented in this paper will require additional
development to deal directly with shell crossing for most
cosmological applications. Alternatively, a good physical understanding
of the nature of the errors introduced by various approximations
may suffice.

In general terms there are three main advantages of the LPT re-expansion
method presented in this paper. (1) The smooth
treatment of the density and velocity fields frees the
results from particle-related shot-noise contamination (see
\citealt*{joyce_towards_2009} for a detailed
discussion of this issue in the context of numerical
simulations). (2) The method can be
fine-tuned to control the numerical error by changing the number of
time steps and/or the Lagrangian order. (3) The scheme
is designed to deal with general initial conditions
including cases where the velocity fields may have a
rotational component. Although this generality has always
been implicit in LPT's formal development 
(\citealt{moutarde_precollapse_1991}; B92), it has mostly been omitted
in applications to real cosmological problems because
vortical modes decay away in expanding cosmologies.
Working in full generality should permit
an LPT based scheme to handle arbitrary initial
data which develops during the course of later phases of
evolution. In addition, given the recent interest in using
information from peculiar velocities to constrain
cosmological parameters, it should prove useful to be
able to handle these modes even in the quasi-linear
regime. 

We envision several applications of our method,
particularly where accurate results are desired on
quasi-linear scales. We mention a few of them below.

One important application is the Baryon Acoustic Oscillation
(BAO) signal reconstruction. The detection of the BAO signal
in the SDSS survey \citep{eisenstein_detection_2005} implies that
measurements of this sort will provide a powerful means to constrain
cosmological parameters. Flows on quasi-linear scales cause
the observed signal to have lower amplitude and peaks
slightly shifted with respect to linear theory
predictions. As observational probes improve, a sub-percent
level precision determination of the BAO scale will require
very accurate reconstruction \citep{percival_baryon_2010}. 
Already techniques based on
single step LPT have been proposed (e.g. \citealt{eisenstein_improving_2007}; \citealt{matsubara_resumming_2008}; \citealt{tassev_towards_2012}).  Our multi-step re-expansion
algorithm can be used to achieve the needed level of error
through control of a combination of time-step and Lagrangian
order. Error control in perturbation techniques is of course
an important element for the establishment of cosmological
constraints \citep*{carlson_critical_2009}.

Another important application is modelling the peculiar
velocity field and investigating the density-velocity
relation. The peculiar velocity information is encoded in
the observed redshift space distortions (RSD) seen in galaxy
surveys. Recent measurements (e.g.
\citealt{guzzo_test_2008}) have demonstrated that this information has the power to
distinguish between various cosmological models.  In linear
theory, the density and velocity divergence are proportional
and consequently, the ratio of the redshift space to real
space power spectra is a constant. Numerical simulations
show a breakdown of that prediction, starting at
scales as large as $k \sim 0.03 h^{-1}$ Mpc \citep*{jennings_modelling_2011}, which would
normally be regarded as well within the linear regime. Understanding
these scales is important. Analytical methods
based on perturbation theory (for e.g., \citealt{bernardeau_quasi-gaussian_1992};
\citealt{kitaura_estimating_2012}) or based on spherical
dynamics (\citealt{bilicki_velocity-density_2008};
\citealt{nadkarni-ghosh_non-linear_2013}) give approximate
answers but definitive answers will require a
more simulation-based approach. LPT re-expansion will be ideal
since it is intrinsically smoother than particle-based approaches. The current
code can track all components of the velocity field enabling
detailed investigations of RSD power spectrum as has
recently been discussed by \citet*{zhang_peculiar_2012}.

Other important applications include the evolution of
non-Gaussianity and the estimation of the growth of voids.
If and when shell crossing is accurately handled the
method of LPT re-expansion may nicely complement existing
methods for modeling the growth of large scale structure.

\section*{acknowledgments}

A part of this work contributed to the Ph. D. thesis of SN
at Cornell University. We thank Ira Wasserman and Rachel
Bean for useful discussions. DFC acknowledges support of
NSF Grant No. AST-0406635 and NASA Grant No. NNG-05GF79G. 
SN acknowledges the hospitality of
the Astronomy department at Cornell University during
subsequent visits. The authors thank Thomas Buchert for useful discussions and pointing out additional references.
\clearpage
\bibliographystyle{mn2e.bst}
\bibliography{mybibtex1,mybibtex2}

\begin{thebibliography}{59}
\expandafter\ifx\csname natexlab\endcsname\relax\def\natexlab#1{#1}\fi

\bibitem[{Adler \& Buchert(1999)}]{adler_lagrangian_1999}
Adler S., Buchert T., 1999, Astronomy and Astrophysics, 343, 317–324

\bibitem[{Bernardeau(1992)}]{bernardeau_quasi-gaussian_1992}
Bernardeau F., 1992, The Astrophysical Journal, 390, L61

\bibitem[{Bilicki \& Chodorowski(2008)}]{bilicki_velocity-density_2008}
Bilicki M., Chodorowski M.~J., 2008, Monthly Notices of the Royal Astronomical
  Society, 391, 1796

\bibitem[{Bouchet {et~al}\mbox{.}(1995)Bouchet, Colombi, Hivon, \&
  Juszkiewicz}]{bouchet_perturbative_1995}
Bouchet F.~R., Colombi S., Hivon E., Juszkiewicz R., 1995, Astronomy and
  Astrophysics, 296, 575

\bibitem[{Bouchet {et~al}\mbox{.}(1992)Bouchet, Juszkiewicz, Colombi, \&
  Pellat}]{bouchet_weakly_1992}
Bouchet F.~R., Juszkiewicz R., Colombi S., Pellat R., 1992, Astrophysical
  Journal, 394, L5–L8

\bibitem[{Buchert(1989)}]{buchert_class_1989}
Buchert T., 1989, Astronomy and Astrophysics, 223, 9–24

\bibitem[{Buchert(1992)}]{buchert_lagrangian_1992}
Buchert T., 1992, Monthly Notices of the Royal Astronomical Society, 254,
  729–737

\bibitem[{Buchert(1994)}]{buchert_lagrangian_1994}
Buchert T., 1994, Monthly Notices of the Royal Astronomical Society, 267, 811

\bibitem[{Buchert {et~al}\mbox{.}(1999)Buchert, Dom\'{i}nguez, \&
  P\'{e}rez-Mercader}]{buchert_extending_1999}
Buchert T., Dom\'{i}nguez A., P\'{e}rez-Mercader J., 1999, Astronomy and
  Astrophysics, 349, 343–353

\bibitem[{Buchert \& Ehlers(1993)}]{buchert_lagrangian_1993}
Buchert T., Ehlers J., 1993, Monthly Notices of the Royal Astronomical Society,
  264, 375–387

\bibitem[{Buchert \& G\"{o}tz(1987)}]{buchert_class_1987}
Buchert T., G\"{o}tz G., 1987, Journal of Mathematical Physics, 28, 2714

\bibitem[{Buchert {et~al}\mbox{.}(1997)Buchert, Karakatsanis, Klaffl, \&
  Schiller}]{buchert_performance_1997}
Buchert T., Karakatsanis G., Klaffl R., Schiller P., 1997, Astronomy and
  Astrophysics, 318, 1–10

\bibitem[{Buchert {et~al}\mbox{.}(1994)Buchert, Melott, \&
  Weiss}]{buchert_testing_1994}
Buchert T., Melott A.~L., Weiss A.~G., 1994, Astronomy and Astrophysics, 288,
  349

\bibitem[{Caldwell {et~al}\mbox{.}(1997)Caldwell, Dave, \&
  Steinhardt}]{caldwell_cosmological_1997}
Caldwell R.~R., Dave R., Steinhardt P.~J., 1997, astro-ph/9708069,
  {Phys.Rev.Lett.80:1582-1585},1998

\bibitem[{Carlson {et~al}\mbox{.}(2009)Carlson, White, \&
  Padmanabhan}]{carlson_critical_2009}
Carlson J., White M., Padmanabhan N., 2009, Physical Review D, 80, 43531

\bibitem[{Catelan(1995)}]{catelan_lagrangian_1995}
Catelan P., 1995, Monthly Notices of the Royal Astronomical Society, 276, 115

\bibitem[{Cooray {et~al}\mbox{.}(2010)Cooray, Holz, \&
  Caldwell}]{cooray_measuring_2010}
Cooray A., Holz D.~E., Caldwell R., 2010, Journal of Cosmology and
  Astro-Particle Physics, 11, 015

\bibitem[{Crocce {et~al}\mbox{.}(2006)Crocce, Pueblas, \&
  Scoccimarro}]{crocce_transients_2006}
Crocce M., Pueblas S., Scoccimarro R., 2006, Monthly Notices of the Royal
  Astronomical Society, 373, 369–381

\bibitem[{Ehlers \& Buchert(1997)}]{ehlers_newtonian_1997}
Ehlers J., Buchert T., 1997, General Relativity and Gravitation, 29, 733–764

\bibitem[{Eisenstein {et~al}\mbox{.}(2007)Eisenstein, Seo, Sirko, \&
  Spergel}]{eisenstein_improving_2007}
Eisenstein D.~J., Seo H.-J., Sirko E., Spergel D.~N., 2007, Astrophysical
  Journal, 664, 675–679

\bibitem[{Eisenstein {et~al}\mbox{.}(2005)Eisenstein, Zehavi, Hogg,
  Scoccimarro, Blanton, Nichol, Scranton, Seo, Tegmark, Zheng, Anderson, Annis,
  Bahcall, Brinkmann, Burles, Castander, Connolly, Csabai, Doi, Fukugita,
  Frieman, Glazebrook, Gunn, Hendry, Hennessy, Ivezić, Kent, Knapp, Lin, Loh,
  Lupton, Margon, {McKay}, Meiksin, Munn, Pope, Richmond, Schlegel, Schneider,
  Shimasaku, Stoughton, Strauss, {SubbaRao}, Szalay, Szapudi, Tucker, Yanny, \&
  York}]{eisenstein_detection_2005}
Eisenstein D.~J. {et~al.}, 2005, The Astrophysical Journal, 633, 560

\bibitem[{Gurbatov {et~al}\mbox{.}(2012)Gurbatov, Saichev, \&
  Shandarin}]{gurbatov_large-scale_2012}
Gurbatov S.~N., Saichev A.~I., Shandarin S.~F., 2012, Physics Uspekhi, 55, 223

\bibitem[{Guzzo {et~al}\mbox{.}(2008)Guzzo, Pierleoni, Meneux, Branchini,
  Le~Fèvre, Marinoni, Garilli, Blaizot, De~Lucia, Pollo, {McCracken}, Bottini,
  Le~Brun, Maccagni, Picat, Scaramella, Scodeggio, Tresse, Vettolani,
  Zanichelli, Adami, Arnouts, Bardelli, Bolzonella, Bongiorno, Cappi, Charlot,
  Ciliegi, Contini, Cucciati, de~la Torre, Dolag, Foucaud, Franzetti,
  Gavignaud, Ilbert, Iovino, Lamareille, Marano, Mazure, Memeo, Merighi,
  Moscardini, Paltani, Pellò, Perez-Montero, Pozzetti, Radovich, Vergani,
  Zamorani, \& Zucca}]{guzzo_test_2008}
Guzzo L. {et~al.}, 2008, Nature, 451, 541

\bibitem[{Jennings {et~al}\mbox{.}(2011)Jennings, Baugh, \&
  Pascoli}]{jennings_modelling_2011}
Jennings E., Baugh C.~M., Pascoli S., 2011, Monthly Notices of the Royal
  Astronomical Society, 410, 2081

\bibitem[{Joyce {et~al}\mbox{.}(2009)Joyce, Marcos, \&
  Baertschiger}]{joyce_towards_2009}
Joyce M., Marcos B., Baertschiger T., 2009, Monthly Notices of the Royal
  Astronomical Society, 394, 751

\bibitem[{Karakatsanis {et~al}\mbox{.}(1997)Karakatsanis, Buchert, \&
  Melott}]{karakatsanis_temporal_1997}
Karakatsanis G., Buchert T., Melott A.~L., 1997, Astronomy and Astrophysics,
  326, 873

\bibitem[{Kasai(1995)}]{kasai_tetrad-based_1995}
Kasai M., 1995, Physical Review D, 52, 5605, copyright {(C)} 2009 The American
  Physical Society; Please report any problems to prola@aps.org

\bibitem[{Kitaura {et~al}\mbox{.}(2012)Kitaura, Angulo, Hoffman, \&
  Gottl\"{o}ber}]{kitaura_estimating_2012}
Kitaura F.-S., Angulo R.~E., Hoffman Y., Gottl\"{o}ber S., 2012, Monthly
  Notices of the Royal Astronomical Society, 425, 2422

\bibitem[{Landau \& Lifschitz(1998)}]{landaulifschitz}
Landau L., Lifschitz E., 1998, Fluid Mechanics, Revised 2nd Edition.
  Butterworth-Heinemann

\bibitem[{Matarrese {et~al}\mbox{.}(1993)Matarrese, Pantano, \&
  Saez}]{matarrese_general-relativistic_1993}
Matarrese S., Pantano O., Saez D., 1993, Physical Review D, 47, 1311–1323

\bibitem[{Matarrese {et~al}\mbox{.}(1994)Matarrese, Pantano, \&
  Saez}]{matarrese_relativistic_1994}
Matarrese S., Pantano O., Saez D., 1994, Monthly Notices of the Royal
  Astronomical Society, 271, 513

\bibitem[{Matarrese \& Terranova(1996)}]{matarrese_post-newtonian_1996}
Matarrese S., Terranova D., 1996, Monthly Notices of the Royal Astronomical
  Society, 283, 400–418

\bibitem[{Matsubara(2008)}]{matsubara_resumming_2008}
Matsubara T., 2008, Physical Review D, 77, 63530

\bibitem[{Melott {et~al}\mbox{.}(1995)Melott, Buchert, \&
  Weiss}]{melott_testing_1995}
Melott A.~L., Buchert T., Weiss A.~G., 1995, Astronomy and Astrophysics, 294,
  345

\bibitem[{Monaco(1997)}]{monaco_lagrangian_1997}
Monaco P., 1997, Monthly Notices of the Royal Astronomical Society, 287,
  753–770

\bibitem[{Monaco {et~al}\mbox{.}(2002)Monaco, Theuns, \&
  Taffoni}]{monaco_pinocchio_2002}
Monaco P., Theuns T., Taffoni G., 2002, Monthly Notices of the Royal
  Astronomical Society, 331, 587–608

\bibitem[{Morita \& Tatekawa(2001)}]{morita_extending_2001}
Morita M., Tatekawa T., 2001, Monthly Notices of the Royal Astronomical
  Society, 328, 815

\bibitem[{Mota {et~al}\mbox{.}(2008)Mota, Shaw, \& Silk}]{mota_magnitude_2008}
Mota D.~F., Shaw D.~J., Silk J., 2008, The Astrophysical Journal, 675, 29–48

\bibitem[{Moutarde {et~al}\mbox{.}(1991)Moutarde, Alimi, Bouchet, Pellat, \&
  Ramani}]{moutarde_precollapse_1991}
Moutarde F., Alimi J.-M., Bouchet F.~R., Pellat R., Ramani A., 1991,
  Astrophysical Journal, 382, 377–381

\bibitem[{Munshi {et~al}\mbox{.}(1994)Munshi, Sahni, \&
  Starobinsky}]{munshi_nonlinear_1994}
Munshi D., Sahni V., Starobinsky A.~A., 1994, The Astrophysical Journal, 436,
  517

\bibitem[{Nadkarni-Ghosh(2013)}]{nadkarni-ghosh_non-linear_2013}
Nadkarni-Ghosh S., 2013, Monthly Notices of the Royal Astronomical Society,
  428, 1166

\bibitem[{Nadkarni-Ghosh \& Chernoff(2011)}]{nadkarni-ghosh_extending_2011}
Nadkarni-Ghosh S., Chernoff D.~F., 2011, Monthly Notices of the Royal
  Astronomical Society, 410, 1454

\bibitem[{Novikov(1969)}]{novikov_nonlinear_1969}
Novikov E.~A., 1969, Soviet Journal of Experimental and Theoretical Physics,
  30, 512

\bibitem[{Peacock(1999)}]{peacock99}
Peacock J., 1999, Cosmological physics. Cambridge University Press

\bibitem[{Percival {et~al}\mbox{.}(2010)Percival, Reid, Eisenstein, Bahcall,
  Budavari, Frieman, Fukugita, Gunn, Ivezić, Knapp, Kron, Loveday, Lupton,
  {McKay}, Meiksin, Nichol, Pope, Schlegel, Schneider, Spergel, Stoughton,
  Strauss, Szalay, Tegmark, Vogeley, Weinberg, York, \&
  Zehavi}]{percival_baryon_2010}
Percival W.~J. {et~al.}, 2010, Monthly Notices of the Royal Astronomical
  Society, 401, 2148

\bibitem[{Press {et~al}\mbox{.}(2002)Press, Teukolsky, Vetterling, \&
  Flannery}]{numrecipes}
Press W., Teukolsky S., Vetterling W., Flannery B., 2002, Numerical Recipes in
  C++. Cambridge University Press

\bibitem[{Rampf(2012)}]{rampf_recursion_2012}
Rampf C., 2012, Journal of Cosmology and Astro-Particle Physics, 12, 004

\bibitem[{Rampf \& Buchert(2012)}]{rampf_lagrangian_2012}
Rampf C., Buchert T., 2012, Journal of Cosmology and Astro-Particle Physics,
  06, 021

\bibitem[{Rampf \& Rigopoulos(2012)}]{rampf_zeldovich_2012}
Rampf C., Rigopoulos G., 2012, {ArXiv} e-prints, 1210, 5446

\bibitem[{Rampf \& Wong(2012)}]{rampf_lagrangian_2012-1}
Rampf C., Wong Y. Y.~Y., 2012, Journal of Cosmology and Astro-Particle Physics,
  06, 018

\bibitem[{Sahni \& Coles(1995)}]{sahni_approximation_1995}
Sahni V., Coles P., 1995, astro-ph/9505005, {Phys.Rept.262:1-135},1995

\bibitem[{Sahni \& Shandarin(1996)}]{sahni_accuracy_1996}
Sahni V., Shandarin S., 1996, Monthly Notices of the Royal Astronomical
  Society, 282, 641–645

\bibitem[{Scoccimarro(1998)}]{scoccimarro_transients_1998}
Scoccimarro R., 1998, Monthly Notices of the Royal Astronomical Society, 299,
  1097–1118

\bibitem[{Scoccimarro \& Sheth(2002)}]{scoccimarro_pthalos:_2002}
Scoccimarro R., Sheth R.~K., 2002, Monthly Notices of the Royal Astronomical
  Society, 329, 629–640

\bibitem[{Tassev \& Zaldarriaga(2012)}]{tassev_towards_2012}
Tassev S., Zaldarriaga M., 2012, Journal of Cosmology and Astro-Particle
  Physics, 10, 006

\bibitem[{Tatekawa(2005)}]{tatekawa_third-order_2005-1}
Tatekawa T., 2005, Physical Review D, 71, 44024

\bibitem[{Tatekawa(2012)}]{tatekawa_fourth-order_2012}
Tatekawa T., 2012, {ArXiv} e-prints, 1210, 8306

\bibitem[{{Zel'dovich}(1970)}]{zeldovich_gravitational_1970}
{Zel'dovich} Y.~B., 1970, Astronomy and Astrophysics, 5, 84

\bibitem[{Zhang {et~al}\mbox{.}(2012)Zhang, Pan, \&
  Zheng}]{zhang_peculiar_2012}
Zhang P., Pan J., Zheng Y., 2012, {ArXiv} e-prints, 1207, 2722

\end{thebibliography}

\appendix
\section{Mathematical Transformations}
\label{mathtrans}
The section below outlines the transformations that are performed on \eqnrefs{diveq} and \eqnrefbare{curleq} to change the dependent variable from ${\bf r}={\bf r}_{EB}$ to ${\bf X}$. For simplicity we drop the subscript `EB'.
The l.h.s. of \eqnref{diveq} is 
\beq
\nabla_r\cdot {\ddot {\bf r}} = \frac{\partial {\ddot r}_i}{\partial r_i} = \frac{\partial {\ddot r_i}}{\partial X_l}\frac{\partial X_l}{\partial r_i}.
\eeq
Einstein's repeated summation convention is followed throughout. The inverse transformation from $X$-space to $r$-space is given as
\beq
\label{xtor}
\frac{\partial X_l}{\partial r_i}= \frac{1}{2 J} \epsilon_{lmn}\epsilon_{ijk}\frac{\partial r_j}
{\partial X_m}\frac{\partial r_k}{\partial X_n},
\eeq
where
\beq
J={\rm Det}\left(\frac{\partial r_i}{\partial X_j}\right)=\epsilon_{abc}\frac{\partial {r_1}}{\partial X_a}\frac{\partial {r_2}}{\partial X_b}\frac{\partial {r_3}}{\partial X_c}=\frac{1}{6}\epsilon_{ipq}\epsilon_{jlm}\frac{\partial  r_i}{\partial  X_j}\frac{\partial {r_p}}{\partial X_l}\frac{\partial { r_q}}{\partial X_m}
\label{jacobdef}
\eeq
and $\epsilon_{ijk}$ is the usual Levi-Civita symbol. 
Define ${\hat L}[{\bf A},{\bf B},{\bf C}] =  \epsilon_{lmq}\epsilon_{ijk} \frac{\partial A_i}{\partial X_l}\frac{\partial B_j}{\partial X_m}\frac{\partial C_k}{\partial X_q}$.  
This gives 
\beq
\nabla_r\cdot {\ddot {\bf r}} =  \frac{1}{2J} \epsilon_{lmn}\epsilon_{ijk} \frac{\partial {\ddot r}_i}{\partial X_l}\frac{\partial r_j}{\partial X_m}\frac{\partial r_k}{\partial X_n} = \frac{1}{2J}{\hat L}[{\ddot {\bf r}},{\bf r},{\bf r}] .
\eeq
The divergence equation then becomes 
\beq
\frac{1}{2J} {\hat L}[{\ddot {\bf r}},{\bf r},{\bf r}]  =
-4\pi G\left(\frac{\rhominit\ainit^3(1+\delta({\bf X},t_0))}{J} + \rhoXinit  (1+3 w) \left(\frac{\ainit}{a}\right)^{3(1+w)}  \right).
\eeq
Using the standard definitions of $\Omega_{\mattersubscript,0}, \Omega_{\darkenergysubscript,0}, \Hinit$, multiplying by $J$ and noting that $J = \frac{1}{6}{\hat L}[{\bf r},{\bf r},{\bf r}]$, the equation is 
\beq
{\hat L}[{\ddot {\bf r}},{\bf r},{\bf r}] =
-3\Hinit^2 \Omega_{\mattersubscript,0}\ainit^3(1+\delta({\bf X},t_0)) -\frac{\Hinit^2}{2}  (1+3 w) \Omega_{\darkenergysubscript,0} \left(\frac{\ainit}{a}\right)^{3(1+w)}   {\hat L}[{\bf r},{\bf r},{\bf r}].
\eeq

The gravitational field is irrotational in Eulerian space so $(\nabla_r \times {\ddot {\bf r}}) = {\bf 0}$. 
Consider the component of equation (\ref{curleq}) along the $i$-th direction.
Again use (\ref {xtor}) to write
\beq(\nabla_r \times {\ddot {\bf r}})_i=  \epsilon_{ijk}\frac{\partial {\ddot r}_k}{\partial r_j}=\epsilon_{ijk}\frac{\partial {\ddot r}_k}{\partial X_l}\frac{\partial X_l}{\partial r_j}=\frac{1}{2J}\epsilon_{ijk}\epsilon_{lmn}\epsilon_{jrs}\frac{\partial \ddot{r}_k}{\partial X_l}\frac{\partial r_r}{\partial X_m}\frac{\partial r_s}{\partial X_n}=0.
\eeq
Using the identity $\epsilon_{ijk}=-\epsilon_{jik}$ and $\epsilon_{jik}\epsilon_{jrs}=\delta_{ir}\delta_{ks}-\delta_{is}\delta_{kr}$ in the above equation 
\beq
-\epsilon_{lmn}\frac{\partial \ddot{r}_k}{\partial X_l}\frac{\partial r_i}{\partial X_m}\frac{\partial r_k}{\partial X_n} + \epsilon_{lmn}\frac{\partial \ddot{r}_k}{\partial X_l}\frac{\partial r_k}{\partial X_m}\frac{\partial r_i}{\partial X_n} = 2  \epsilon_{lmn}\frac{\partial \ddot{r}_k}{\partial X_l}\frac{\partial r_k}{\partial X_m}\frac{\partial r_i}{\partial X_n} =0.
\eeq
Here we have set each component of the vector $\nabla_r\times{\ddot {\bf r} }$ in the $r$ basis equal to zero. One can also express the components of $\nabla_r\times{\ddot {\bf r} }$ in the $X$ basis where the two bases are related by 
$\hat r_i=\frac{\partial X_p}{\partial r_i} \hat X_p$. This gives 
\beq
 \epsilon_{lmn}\frac{\partial \ddot{r}_k}{\partial X_l}\frac{\partial r_k}{\partial X_m}\frac{\partial r_i}{\partial X_n} \frac{\partial X_p}{\partial r_i} \hat X_p  = 0.
 \eeq 
 But $\partial r_i/\partial X_n \cdot \partial X_p/\partial r_i = \delta_{pn}$. Since the basis vectors are all independent the individual components must be zero. The simplified condition for each $n$ then becomes
\beq
\label{mainT2}
\epsilon_{nlm}\frac{\partial \ddot{r}_k}{\partial X_l}\frac{\partial r_k}{\partial X_m}=0. 
\eeq
Defining ${\hat T}_q [{\bf A},{\bf B}] = \epsilon_{lmq}\frac{\partial A_k}{\partial X_l}\frac{\partial B_k}{\partial X_m}$ it is rewritten as ${\hat {\bf T}}[{\ddot {\bf r}},{\bf r}] = 0$. 
${\hat L}$ and ${\bf \hat T}$ are linear in each of their variables. 

\section{Computational details of the EB scheme}
\label{app:initialcond}

In the EB frame, the equations to be solved have the following form 
\bea
D_t^L \left[ \nabla_{\rm X}\cdot{\bf p}^{(1)} \right] &=&-\frac{3}{2}\Hinit^2\Omega_{\mattersubscript,0}\ainit^3\delta({\bf X},t_0)
\label{firstorderL2}\\
D_t^T \left[\nabla_{\rm X} \times {\bf p}^{(1)}\right]&=&0 
\label{firstorderT2} \\
D_t^L\left[ \nabla_{\rm X}\cdot{\bf p}^{(n)} \right] &=& S^{(n,L)}
\label{higherorderL2}\\
D_t^T \left[ \nabla_{\rm X} \times {\bf p}^{(n)}\right] &= & {\bf S}^{(n,T)},
\label{higherorderT2}
\eea 
where $n$ refers to orders higher than the first and 
\bea
\label{DtLapp}
 D_t^L &=& \left(2a \ddot a+\frac{3}{2}a^2 \Hinit^2 (1+3 w) \Omega_{\darkenergysubscript,0} \left(\frac{\ainit}{a}\right)^{3(1+w)} + a^2 \frac{d^2}{dt^2}\right)\\
\label{DtTapp}
D_t^T &= &\left(-\ddot a + a \frac{d^2}{dt^2}\right).
\eea
The explicit form for the source terms is
\bea
\nonumber S^{(n,L)}
&=&  \mathop{\sum_{\alpha, \beta}}_{\alpha + \beta = n} \left(-\frac{1}{2} \ddot a  - \frac{3}{4} a \Hinit^2 (1+3 w) \Omega_{\darkenergysubscript,0} \left(\frac{\ainit}{a}\right)^{3(1+w)} \right) {\hat L}[{\bf p}^{(\alpha)},  {\bf p}^{(\beta)}, {\bf X}] \\
\nonumber & &  -\mathop{\sum_{\alpha, \beta}}_{\alpha + \beta = n} a {\hat L}[{\ddot {\bf  p}}^{(\alpha)}, {\bf p}^{(\beta)}, {\bf X}] - \mathop{\sum_{\alpha, \beta, \gamma}}_{\alpha + \beta + \gamma = n} \frac{1}{2} 
{\hat L}[{\ddot {\bf  p}}^{(\alpha)},  {\bf p}^{(\beta)}, {\bf p}^{(\gamma )}] \\
&& - \mathop{\sum_{\alpha, \beta, \gamma}}_{\alpha + \beta + \gamma = n}
\frac{1}{4}\Hinit^2 (1+3 w) \Omega_{\darkenergysubscript,0} \left(\frac{\ainit}{a}\right)^{3(1+w)} {\hat L}[{\bf  p}^{(\alpha)},  {\bf p}^{(\beta)}, {\bf p}^{(\gamma )}] \\ 
{\bf S}^{(n,T)}
&=&
- \mathop{\sum_{\alpha, \beta}}_{\alpha + \beta = n}{\hat {\bf T}}[{\ddot {\bf  p}}^{(\alpha)},{\bf p}^{(\beta)}],
\eea
where $\alpha, \beta,\gamma$ can take any values from $1$ to $n-1$ and they add up to $n$. These are subject to initial conditions 
\bea
\label{initialorder1} \mbox{ {\it at first order}} \; \; \; {\bf p}^{(1,L/T)}({\bf X},t_0) &=&0,    \; \;  {\dot {\bf p}}^{(1,L/T)}({\bf X},t_0) = {\bf v}^{L/T}({\bf X}, t_0) \\
\mbox{ {\it at higher order }} {\bf p}^{(n, L/T)}({\bf X},t_0) &=& 0,   \label{initialpn2} \; \; {\dot {\bf p}}^{(n, L/T)}({\bf X},t_0) =0.
\label{initialordern}
\eea

Since the spatial and temporal operators in \eqnrefs{firstorderL2} - \eqnrefbare{higherorderT2} commute, the displacement field can be written as a linear combination spatial vectors with purely time dependent coefficients
\beq {\bf p}^{(n,L/T)} = \sum_{i=1}^{Z_{n,L/T}} b_i^{(n,L/T)} {\bf  F}_i^{(n,L/T)},
\label{pexp1}
\eeq 
where $b$ and ${\bf F}$ are functions of $t$ and ${\bf X}$ respectively i.e., $b\equiv b(t)$ and ${\bf F} \equiv {\bf F}({\bf X})$ and 
$Z_{n,L}$ ($Z_{n,T}$) denote the number of independent longitudinal (transverse) terms at the $n$-th order. The superscripts denote the order and type of the term. 
Using this decomposition, one can denote the higher order source terms as 
\bea
 S^{(n,L)} &=& \sum_{i=1}^{Z_{n,L}} h^{(n,L)}_i(t) \cdot {\hat L}_i[{\bf F}^\alpha, {\bf F}^\beta,{\bf F}^{\gamma}]\\
{\bf S}^{(n,T)} &=&  \sum_{i=1}^{Z_{n,T}} h^{(n,T)}_i (t)\cdot{\hat {\bf T}}_i[{\bf F}^\alpha,{\bf F}^{\beta}]
\eea
respectively, where the superscripts of ${\bf F}$s take any value between $0$ to $n-1$ (0 corresponding to ${\bf F}({\bf X})={\bf X}$) and add up to $n$. It is possible to calculate all the independent spatial source terms symbolically at all orders using the symmetries and properties of the ${\hat L}$ and ${\hat {\bf T}}$ from which the numbers $Z_{n,L}$ and $Z_{n,T}$ can be determined and temporal source functions $h(t)$ can be extracted.

The initial perturbation is described by one scalar field $\delta({\bf X},t_0)$ corresponding to the initial density field in the EB frame and two vector fields ${\bf v}^L({\bf X}, t_0)$ and ${\bf v}^T({\bf X}, t_0)$ corresponding respectively to the curl-free and divergence-less parts of the initial velocity field (in the EB frame). The initial acceleration vector can be constructed from the initial density field via Poisson's equation and is also curl-free. This gives $Z_{1,L} =2$ and $Z_{1,T} = 1$. Table \ref{tableZ} shows the number of terms at various orders. If the initial conditions start with zero transverse velocity then $Z_{1,T} =0$. In this special case the total number of equations to solve reduces from 9 to 4 at second order and from 64 to 20 at third order.

Substituting \eqnref{pexp1} up to first order in \eqnrefs{firstorderL2} and \eqnrefbare{firstorderT2} gives,
\bea
D_t^L b_1^{(1,L)}[\nabla_X \cdot{\bf F}_1^{(1,L)}]
+ D_t^L b_2^{(1,L)}[\nabla_X \cdot{\bf F}_2^{(1,L)}]&=& -\frac{3}{2}\Hinit^2\Omega_{\mattersubscript,0}\ainit^3\delta({\bf X},t_0), \\
D_t^T b_1^{(1,T)}[\nabla_X \times {\bf F}_1^{(1,T)} ]&=&0.
\eea
The initial conditions that ${\bf p}^{(1)}$ satisfies are given by \eqnrefs{initialorder1}. There is a choice to be made in how these initial conditions translate to conditions on the temporal and spatial functions. We choose to set them as shown below. 

\begin{flushleft}
\begin{tabular}{l|l|l}
 $\nabla_X \cdot{\bf F}_1^{(1,L)} = \delta({\bf X},t_0) $ & ${\bf F}_2^{(1,L)} =   {\bf v}^{L}({\bf X}, t_0)$ & ${\bf F}_1^{(1,T)}  =   {\bf v}^{T}({\bf X}, t_0)$ \\  
$D_t^L b_1^{(1,L)}=  -\frac{3}{2}\Hinit^2\Omega_{\mattersubscript,0}\ainit^3$ & $D_t^L b_2^{(1,L)} = 0$ &$D_t^T b_1^{(1,T)} = 0. $\\
$b_{1}^{(1,L)}(t_0) =0$ &$ b_2^{(1,L)}(t_0) =0 $&$  b_1^{(1,T)}(t_0) =0$\\
$ {\dot b}_1^{(1,L)}(t_0) =0 $&$ {\dot b}_2^{(1,L)}(t_0)=1$ &$ {\dot b}_1^{(1,T)}(t_0)=1$
\label{Fandb1}
\end{tabular}
\end{flushleft}
Substituting \eqnref{pexp1} in the higher order equations \eqnrefs{higherorderL2} and \eqnrefbare{higherorderT2} and imposing \eqnref{initialordern} gives 
\begin{flushleft}
\begin{tabular}{l|l}
$ \nabla_X \cdot {\bf F}_i^{(n,L)} =  {\hat L}_i[{\bf F}^\alpha, {\bf F}^\beta,{\bf F}^\gamma] $ & $ \nabla_X \times {\bf F}_i^{(n,T)} =  {\hat {\bf T}}_i[{\bf F}^\alpha, {\bf F}^{\beta}] $.\\
$ D_t^L b_i^{(n,L)}(t) =  h^{(n,L)}_i(t)$& $ \nonumber D_t^T b_i^{(n,T)}(t) = h^{(n,T)}_i(t)$. \\
$b_i^{(n,L)}(t_0) = 0$ & $ b_i^{(n,T)}(t_0) = 0$.\\
${\dot b}_i^{(n,L)}(t_0) = 0$ & $ {\dot b}_i^{(n,T)}(t_0) = 0$.
\end{tabular}
\end{flushleft}
All the spatial solutions are subject to periodic boundary conditions and are solved using Fourier transforms. 

\begin{center}
\ra{1.4}
\begin{table}
\caption{Number of transverse and longitudinal terms as a function of Lagrangian order and type of initial conditions.}
\begin{tabular}{|c|c|c|c|}
\hline
Order& $\mbox{ Zel'dovich initial conditions} $ & ${\bf v}^T = 0 $& ${\bf v}^T\neq 0 $ \\
\hline
n=1& $Z_L=2,Z_T =0$& $Z_L=2,Z_T =0$ & $Z_L=2,Z_T =1$\\
n=2 &$Z_L=3,Z_T =0$& $Z_L=3 ,Z_T =1$ & $Z_L=6,Z_T =3$\\
n=3 &$Z_L=10,Z_T =6$& $Z_L=12 ,Z_T =8$ & $Z_L=37,Z_T =27$\\
\hline
\end{tabular}
\label{tableZ}
\end{table}
\end{center}

\section{Formal requirements of the EB solution}
 \label{zeromean}
 
The formal derivation by Ehlers and Buchert (EB97) demands that the displacement in the EB frame satisfies 
\beq
\int_V {\bf p}({\bf X},t)d^3 X = 0.
\label{pzerocond}
\eeq
for all times $t$. At the initial time the definition of the Lagrangian coordinate implies ${\bf p}({\bf X},t_0) =0$ and \eqnref{pzerocond} is trivially satisfied. We want to prove that if the initial data satisfies 
\bea 
\langle \delta({\bf X}, t_0) \rangle_X &=& 0\\
\langle {\bf v}({\bf X}, t_0) \rangle_X &=& 0
\eea
and the system is periodic, then \eqnref{pzerocond} holds at all times. We refer to these as the `zero-mean' initial conditions. \\

{\it Proof}: The displacement ${\bf p}({\bf X}, t)$ is a time-dependent linear combination of some spatial functions ${\bf F}({\bf X})$. 
We will prove that each ${\bf F}({\bf X})$ satisfies $\langle {\bf F}({\bf X}) \rangle_X =0 $ and therefore $\langle {\bf p}({\bf X},t) \rangle_X =0$. 

At first order, the spatial function is either a solution of $\nabla_X \cdot {\bf F}({\bf X}) = \delta({\bf X}, t_0)$ or  ${\bf F}({\bf X}) = {\bf v}({\bf X}, t_0)$. Since the initial density integrates to zero over the volume and the system is periodic, the solution also has the same property. Therefore, at first order all spatial ${\bf F}$s has mean zero. 
  
At higher orders the ${\bf F}$s are periodic solutions of $\nabla_X \cdot {\bf F} = S^L({\bf X})$ or $\nabla_X \times {\bf F} = {\bf S}^T({\bf X})$, where $S^L, {\bf S}^T$ are combinations of lower order ${\bf F}$s. So if $S^L({\bf X})$ and ${\bf S}^T({\bf X})$ have zero mean then periodicity will imply that each higher order ${\bf F}$ has mean zero. Here we present the proof only for the longitudinal source terms $S^L({\bf X}) = {\hat L}[{\bf F}^\alpha, {\bf F}^\beta, {\bf F}^\gamma]$. The proof for the transverse sources ${\hat {\bf T}}[{\bf F}^\alpha, {\bf F}^\beta]$ follows a similar calculation. 

Let $I = \langle {\hat L}[{\bf F}^\alpha, {\bf F}^\beta, {\bf F}^\gamma] \rangle $. To keep in mind the periodic nature of the system, we refer to space as the 3-torus ${\bf T}^3$. 
Write 
$I=\frac{1}{V} \epsilon_{ijk}I'_{ijk}$, where
\beq I'_{ijk}=\epsilon_{pqr}
\int_{{\bf T}^3} d^3X \frac{\partial F^\alpha_i}{\partial X_p}\frac{\partial F^\beta_j}{\partial X_q}\frac{\partial F^\gamma_k}{\partial X_r}. \eeq 
Here the spatial functions ${\bf F}$ can either be periodic functions of ${\bf X}$ or equal to ${\bf X}$. These are higher order source terms. Hence at least one of ${\bf F}$s is not ${\bf X}$. Without loss of generality let this correspond to the term $\partial F_k/\partial X_r$. Integrating by parts over $X_r$ and using periodicity gives  
\beq
I'_{ijk} =-\epsilon_{pqr}\int_{{\bf T}^3} d^3X \frac{\partial}{\partial X_r}\left(\frac{\partial F^\alpha_i}{\partial X_p}\frac{\partial F^\beta_j}{\partial X_q}\right) F_k^\gamma = 
-\epsilon_{pqr}\int_{{\bf T}^3} d^3X  \left[\left(\frac{\partial^2 F^\alpha_i}{\partial X_r\partial X_p}\right)\frac{\partial F^\beta_j}{\partial X_q}  +  \left(\frac{\partial^2 F^\beta_j}{\partial X_r\partial X_q}\right)\frac{\partial F^\alpha_i}{\partial X_p}   \right]F_k^\gamma. 
\eeq
The first integrand is symmetric under the exchange $r \leftrightarrow p$ and the second under the exchange $r\leftrightarrow q$. $\epsilon_{pqr}$ is antisymmetric under these exchanges. Therefore, $I'_{ijk}=0$ and hence $I=0$. 
Similarly one can show that the source terms for solving the transverse vectors also integrate to zero. Thus \eqnref{pzerocond} holds for all $t$.

\section{Setting the initial conditions along the Zel'dovich curve}
\label{App:zeldovich}
It is generally common in cosmology to assume that the initial velocity is proportional to the initial acceleration \citep{zeldovich_gravitational_1970}; the proportionality constant is set by requiring that there be no perturbations at the big bang. This is also equivalent to having no decaying modes i.e. no negative powers of $t$ in the displacement. At very early times (recombination) the universe is purely matter dominated with $\Omega=1$ and the first order displacement evolves as (B92) 
\beq
{\bf p}^{(1)} = \left (\frac{t}{t_0} \right)^{-1/3}  \left[ -\frac{2 \ainit}{5}  {\bf F}_1^{(1,L)} - \frac{3 t_0}{5}{\bf v}^L \right]
 +   \left (\frac{t}{t_0} \right)^{4/3} \left[ -\frac{3\ainit}{5}  {\bf F}_1^{(1,L)} +  \frac{3 t_0}{5}{\bf v}^L \right]
 +   \ainit \left (\frac{t}{t_0} \right)^{2/3} {\bf F}_1^{(1,L)}.
\eeq
Setting the coefficient of the negative power of $t$ to zero gives 
\beq 
{\bf v}^{L}  = - \frac{2}{3} \cdot\frac{\ainit}{t_0} {\bf F}_1^{(1,L)} = -{\dot a}(t_0)  {\bf F}_1^{(1,L)}. 
\label{linearzel}
\eeq
Note that this does not guarantee that there are no decaying modes from the higher order solution. The terms $(t/t_0)^{-1/3}$ and $(t/t_0)^{4/3}$ arise from the homogeneous part of the solution and each higher order solution will contain them since the temporal derivative operator is the same at all orders. An alternate way is to choose the initial velocity such that the scaled velocity divergence $\delta v = (3 \Hinit)^{-1}\nabla_{r_{OBS}} \cdot {\bf v}_{OBS}$ and $\delta({\bf X},t_0)$ at each point satisfy the non-linear Zel'dovich relation based on the top-hat analysis (NC11). In this case the relationship in \eqnref{linearzel} is not satisfied; decaying modes will be present even at first order. The former prescription is used more often in literature and in this paper. We also checked that the difference in the two ways of setting the initial conditions is very small ($\sim  10^{-8}$) when the starting epoch is $z \sim 1000$ (recombination). 

\section{1D perturbations}
\label{App:1D}

This appendix proves the results for the 1-D case stated in \S \ref{sec:1dexample}. 

\subsection{Frame shift for a single step}
Since the problem is essentially 1-d, we will denote all spatial functions as scalar functions. The Lagrangian coordinate in the observer (EB) frame is denoted as $Y$ $(X)$; $Y=X$. The initial density and velocity in the observer frame are $\delta_{OBS}(Y,t_0)$ and $v_{OBS}(Y,t_0)$ respectively. 
The initial data in the EB frame is 
\beq
\delta_{EB}(X,t_0) = \delta_{OBS}(Y,t_0);  \; \; v_{EB}(X,t_0) = v_{OBS}(Y,t_0) -v_{c,0},
\eeq
where $v_{c,0} =  \langle v_{OBS}(Y,t_0) \rangle $. The acceleration in the observer's frame is given by $F_\delta(Y,t_0)$ where $d F_\delta/dY =\delta_{OBS}(Y,t_0)$. The densities are the same, hence $F_\delta(X,t_0) = F_\delta(Y,t_0)$. Hereafter we drop the subscripts on the density since it is the same in both frames. 

The displacement in the EB frame at any later time is given as 
\beq 
p(X,t) = b_1(t) F_1(X) + b_2(t) F_2(X), 
\label{pEB}
\eeq
where $ F_1(X) = F_\delta(X,t_0)$ and $F_2(X) =  v_{EB}(X,t_0)$. 
The physical coordinate 
\beq
r_{EB}(X,t) =  a X + p(X,t)
\label{comov}
\eeq and the corresponding Jacobian is 
\beq 
 J(X,t)= a^3 \left(1+ \frac{1}{a}\frac{dp}{dX}\right).
\label{jacob}
\eeq 
The comoving coordinate is $x_{EB} = r_{EB}/a$. Restricting to 1D and substituting \eqnref{comov} and \eqnrefbare{jacob} in \eqnref{eqforshift}, the frameshift equation is  
 \beq 
\frac{d}{dt} \left(a^2 \frac{d \Delta x}{dt} \right) = -\frac{1}{L} \int (a {\ddot p} - {\ddot a} p) \left(1 + \frac{1}{a}\frac{d p}{dX} \right) dX.
\eeq
Since $p$ is a periodic function of $X$, this gives 
\beq 
\frac{d}{dt} \left(a^2 \frac{d \Delta x}{dt} \right) = -\frac{1}{L} \int {\ddot p} \frac{d p}{dX}  dX.
\label{shiftpEB}
\eeq
Substituting for $p$ from \eqnref{pEB}
\beq 
\frac{d}{dt} \left(a^2 \frac{d \Delta x}{dt} \right) = -\frac{1}{L} \int ({\ddot b_1} b_2 F_1 F'_2 + {\ddot b_2} b_1 F_2 F'_1) dX,
\label{eqnEB}
\eeq 
where prime denotes differentiation w.r.t. $X$ or $Y$. 
Using the equations for ${\ddot b}_1$ and ${\ddot b}_2$ given in Appendix \ref{app:initialcond}, applying the product rule for derivatives and using periodicity of the spatial functions gives 
\beq 
\frac{d}{dt} \left(a^2 \frac{d \Delta x}{dt} \right) = - \left(3 {\ddot a} +  \frac{3}{2} (1+3w) a \Omega_{X,i} \left(\frac{a_i}{a}\right)^{3(1+w)}  \right)b_2  \frac{1}{L} \int (F_1 F'_2 ) dX \propto \int F_\delta(Y,t_0) v_{OBS}'(Y,t_0) dY, 
\label{eqnEB2}
\eeq 
with initial conditions $\Delta x(t_0)=0$ and ${\dot \Delta}x(t_0) = v_{c,0}/a_0$.
Consider the case where $v_{OBS}(Y,t_0) \propto F_\delta(Y,t_0)$. The condition that the mean density be zero implies $\langle F_\delta(Y,t_0) \rangle = v_{c,0} =0$. Thus the initial conditions as well as source term of \eqnref{eqnEB2} are zero and hence $\Delta x=0$ for all $t$. Thus, single step LPT with Zel'dovich initial conditions requires no shifts. 

\subsection{Maintaining the Zel'dovich condition}
Let the initial time at the start at the first step be $t_0$ and consider taking a second step at time $t_1$. 
Let $Y_0$, $Y_1$ be the Lagrangian coordinates in the observer frame at time $t_0$, $t_1$ respectively. $Y_1$ is related to $Y_0$ as 
\beq 
Y_1 = Y_0+ a^{-1} p(X,t_1) + \Delta x(t_1)
\label{Y1toX}
\eeq
If $v_{OBS}(Y_0,t_0) \propto F_\delta(Y_0,t_0) $, then $v_{c,0} =0$ and $v_{EB}(X,t_0) = v_{OBS}(Y_0,t_0)$. Combining with \eqnref{pEB}, this gives 
\beq p(X,t_1) \propto F_\delta(Y_0, t_0) \; \; \; {\rm and} \; \; \;  {\dot p}(X,t_1) \propto F_\delta(Y_0, t_0).
\label{peb}
\eeq
The velocity in the observer frame at the start of the second step is
\beq
v_{OBS}(Y_1,t_1) = {\dot p}(X,t_1) - \frac{{\dot a}(t_1)}{a(t_1)} p(X,t_1) + a {\dot \Delta}x(t_1). 
\label{vobs}
\eeq
The acceleration at the start of the second step is
\beq 
F_\delta(Y_1,t_1) = \int \delta(Y_1,t_1) dY_1  = \frac{(1+ \delta(Y_0,t_0))a^3}{J(X,t)} -1,  
\eeq
where $J(X,t)$ is given by \eqnref{jacob}. 
Transforming from $Y_1$ to $X(=Y_0)$ using \eqnref{Y1toX} 
\bea 
F_\delta(Y_1,t_1) &=& \int(1+ \delta(Y_0,t_0) )dY_0 - \int\left(1+\frac{1}{a} \frac{dp}{dX} \right) dX\\
&=& F_\delta(Y_0,t_0) -\frac{p(X,t_1)}{a}.
\eea
When the initial acceleration and velocity are proportional, the above equation can be combined with \eqnref{peb} to give $F_\delta(Y_1, t_1) \propto F_\delta(Y_0,t_0) $. Since the frame shifts are zero, \eqnref{peb} and \eqnrefbare{vobs} give $v_{OBS}(Y_1,t_1)  \propto F_\delta(Y_0, t_0)$. This proves that if the initial acceleration and velocity are proportional they remain proportional at all subsequent times. 

\subsection{Conservation of mean velocity}
Using the periodicity of $p(X,t)$, \eqnref{shiftpEB} can be re-written as 
\beq 
\frac{d}{dt} \left(a^2 \frac{d \Delta x}{dt} \right) = \frac{d}{dt} \left(-\frac{1}{L} \int {\dot p} \frac{d p}{dX}  dX\right).
\eeq
This gives 
\beq 
{\dot \Delta} x(t) =   \frac{v_{c,0} a_0}{a^2} - \frac{1}{a^2} \left(\frac{1}{L} \int {\dot p} \frac{d p}{dX}  dX\right) . 
\label{shift}
\eeq 
The velocity in the observer frame at any time is given by \eqnref{vobs}.
The mean at any time $t$ is 
\bea
\langle v_{OBS}(Y_1,t) \rangle &=&  \frac{1}{L} \int v_{OBS}(Y_1,t) dY_1 \\
&=&   \frac{1}{L} \int \left({\dot p}- \frac{{\dot a}}{a} p + a {\dot \Delta}x  \right) \left(1+ \frac{1}{a} \frac{dp}{dX} \right) dX\\
&=& a {\dot \Delta}x   + \frac{1}{a} \left( \frac{1}{L} \int {\dot p} \frac{d p}{dX}  dX \right).
\eea
Substituting for ${\dot \Delta}x$ from \eqnref{shift}, 
\beq \langle v_{OBS}(Y_1,t) \rangle = \frac{v_{c,0} a_0}{a}.\eeq
Thus, for a 1-d system the mean velocity, like the mean momentum, is conserved modulo a `$1/a$ decay' due to the Hubble drag. 

\section{Spherical top-hat }
\label{app:SPT}
This section describes the details involved in setting up the compensated top-hat configuration for section \ref{sec:gridNs}.

The exact compensated top-hat function consists of a spherical overdense region surrounded by a compensating underdense vacuum region. 
Let $a$ and $b$ be the radii of the overdense and compensating regions respectively.
The initial density profile is given by 
\beq
\delta({\bf X},t_0)=
\left\{
\begin{array}{lr}
 \delta_0 &0 \leq |{\bf X}| <a\\
-1         & a \leq |{\bf X}| < b\\
0 & |{\bf X}| \geq b
\end{array}
\right.
\label{eq:rhoi}
\eeq
We chose $\delta_0 =10$, $a=1/4$ and $b=11^{1/3}/4$. The box length $L_{box}$ is chosen to be two units in length centred around the point $(1/10,-1/11, 1/(2 \pi))$. The choice of parameters ensures that the entire profile is well represented within the box and the offset ensures that no special symmetry is exploited in the test. The profile is discontinuous at $X=a$ and $X=b$. The Fourier transform of a discontinuous function has power at all wave numbers and the Gibbs phenomenon prevents such functions from being completely represented by Fourier transforms on any finite size grid. The initial profile is hence smoothed with a Gaussian filter of width $\sigma$ to give 
 \beq 
\delta_{\sigma}({\bf X},t_0)=\int_0^{L_{box}} \delta({\bf X},t_0)\cdot \frac{\exp(-X^2/ 2\pi \sigma^2)}{(2\pi\sigma^2)^{3/2}}d^3X. 
\eeq  
The smoothing is performed analytically in real space and the resulting function is numerically evaluated on the grid. To ensure periodicity of the initial
data, contribution of the 26 nearest neighbours cells is added. The contribution of cells beyond the nearest neighbours was zero to machine precision.
For the density profile in \S \ref{sec:gridNs}, the smoothing parameter is $\sigma=1/12$.

\section{Generation of the transverse velocity}
\label{app:vTgenerate}
The Eulerian vorticity in the observer frame is $\nabla_{r_{OBS}} \times {\bf v}_{OBS}$. 
\beq
\nabla_{r_{OBS}}  \times {\bf v}_{OBS} = \epsilon_{ijk} \frac{\partial v_{OBS,k}}{\partial r_{OBS,j}}.
 \eeq
LPT expresses ${\bf v}_{OBS}$ as functions of the Lagrangian variable ${\bf Y}$ and not ${\bf r}_{OBS}$. 
Using the relations in Appendix \ref{mathtrans} and properties of the Levi-Civita tensor gives
\bea
(\nabla_{r_{OBS}}  \times {\bf v}_{OBS})_i &=&\frac{1}{2J}  \epsilon_{ijk}\epsilon_{jpq}\epsilon_{lmn}  r_{p,m} r_{q,n} v_{k,l}\\
\nonumber &=& \frac{1}{2J} (\delta_{kp} \delta_{iq} - \delta_{ip}\delta_{kq} )  \epsilon_{lmn}  r_{p,m} r_{q,n} v_{k,l}\\
\label{Euleriancurl}&=& \frac{1}{2J} \epsilon_{lmn} (r_{k,m} r_{i,n} v_{k,l} - r_{i,m} r_{k,n} v_{k,l}   ) \\ 
&=& \frac{1}{J} \epsilon_{mln} ( r_{i,m} r_{k,n} v_{k,l})
\eea
where the `comma' denotes differentiation with respect to the Lagrangian variable ${\bf Y} = {\bf X}$. Substituting for ${\bf r}_{OBS}$ and ${\bf v}_{OBS}$ from \eqnrefs{robs} and \eqnrefbare{vobsofY} gives rise to three types of terms in the expression: 
$\{ \epsilon_{ilk}{\dot p}_{k,l}, \epsilon_{ilk}p_{k,l} \}, \{\epsilon_{iln} p_{k,n} {\dot p}_{k,l}\}$ and $\{ \epsilon_{lmn} p_{i,m} p_{k,n} {\dot p}_{k,l}\}$. 
The first kind corresponds to the vorticity in the Lagrangian frame while the remaining two have no obvious physical interpretation. In general, all three types of terms are non-zero. For a first order LPT calculation (i.e., the displacement is accurate to first order in the expansion parameter and leading errors are second order), ${\dot {\bf p}}$ is always proportional to ${\bf p}$ and the anti-symmetry of the Levi-Civita tensor ensures that all three types of terms vanish. The second order LPT series does not have this feature. In particular, the second order displacement term in the series ${\bf p}^{(2)}$ is not proportional to the first order ${\bf p}^{(1)}$ and hence combinations of these (which are third order) will in general not be zero. Thus, the Eulerian vorticity will in general have non-zero contributions which are third order and higher (up to order 6) in the expansion parameter. 
Thus, when calculations are performed with second order LPT, the Eulerian vorticity is non-zero although the Lagrangian vorticity is zero. 

\end{document}